\documentclass[prb,twocolumn,showpacs,aps,floats,floatfix,superscriptaddress]{revtex4}
\usepackage{epsfig}
\usepackage{colordvi}
\usepackage{graphicx}
\usepackage{amssymb}

\begin{document}
\title{On the combined use of GW approximation and cumulant expansion in the calculations of quasiparticle spectra: The paradigm of Si valence bands}
\author{Branko Gumhalter\footnote{Corresponding author. Email: branko@ifs.hr}}
\affiliation{Institute of Physics, HR-10000 Zagreb, Croatia}
\author{Vjekoslav Kova\v{c}}
\affiliation{Department of Mathematics, University of Zagreb, HR-10000 Zagreb, Croatia}
\author{Fabio Caruso}
\affiliation{Department of Materials, University of Oxford, Parks Road, Oxford OX1 3PH, United Kingdom}
\author{Henry Lambert}
\affiliation{Department of Materials, University of Oxford, Parks Road, Oxford OX1 3PH, United Kingdom}
\author{Feliciano Giustino}
\affiliation{Department of Materials, University of Oxford, Parks Road, Oxford OX1 3PH, United Kingdom}

\date{\today}

\begin{abstract}
Since the earliest implementations of the various GW approximations and cumulant expansion in the calculations of quasiparticle propagators and spectra, several attempts have been made to combine the advantageous properties and results of these two theoretical approaches. While the GW plus cumulant approach has proven successful in interpreting photoemission spectroscopy data in solids, the formal connection between the two methods has not been investigated in detail. By introducing a general bijective integral representation of the cumulants, we can rigorously identify at which point these two approximations can be connected for the paradigmatic model of quasiparticle interaction with the dielectric response of the system that has been extensively exploited in recent interpretations of the satellite structures in  photoelectron spectra. We establish a protocol for consistent practical implementation of the thus established GW+cumulant scheme, and illustrate it by comprehensive state-of-the-art first-principles calculations of intrinsic angle-resolved photoemission spectra from Si valence bands.
\end{abstract}


\maketitle

\newcommand{\bq}{\begin{equation}}
\newcommand{\eq}{\end{equation}}

\newcommand{\barr}{\begin{eqnarray}}
\newcommand{\earr}{\end{eqnarray}}

\section{Introduction}
\label{sec:intro}

In the past two decades a number of attempts has been made to combine the algorithms developed for calculating the band structure and quasiparticle properties in solids within the so-called GW approximations (GWA) for the quasiparticle self-energy [\onlinecite{Hedin,BLundqvistGW,Hybertsen1986,Hybertsen1987,Eguiluz,MahanGW,AryasetiawanGunnarsson,Aulbur1999,Hedin1999,Farid1999,Onida,MariniSoleRubio}] with the powerful cumulant expansion or linked cluster approach [\onlinecite{Kubo,Brout,ND,MHRT,Mahan_book,Langreth,LangrethNobel}] used for treating multiple excitation processes in the coupled quasiparticle-heatbath systems [\onlinecite{Mahan1966,Doniach,ChangLangreth,Dunn,HedinPS,Skinner,Hsu_Skinner,AlmbladhHedin,Meden,Bechstedt,AHK,Inglesfield,4cumul,Lazic_PRL,Lazic_PRB,Reining,PSS2012,Louie,Pavlyukh,Kas,Lischner2DEG,Feliciano,Fadley,Caruso,SGLD}]. The major rationale for  combining these two approaches in the treatment of spectral properties of multiply excited systems stems from the empirically verified property of GWA to describe well the quasiparticle spectrum at the excitation threshold, on the one hand, and of cumulant expansion to reliably reproduce the multiexcitation structures where GWA fails, on the other hand. Here the relevant comparisons and assessments of the performance of two methods are made for the same generic interaction. In the GWA this is the dynamically screened electronic interaction $W$ calculated in the the linear response. The ensuing cumulant approach is generated by the dynamic component of the same interaction which has the properties of a boson propagator. Thereby the problem maps onto the treatment of the coupled quasiparticle-boson system. The transition from all-electron to quasiparticle-boson model in the context of cumulant approach was discussed in Refs. [\onlinecite{Langreth,LangrethNobel,AlmbladhHedin}].      

 In the context of {\it ab initio} calculations cumulant expansion
was first employed in combination with the GW approximation
to study the satellites in the photoemission spectra of simple metals [\onlinecite{AHK}].
Subsequent studies revealed that this approach may also prove useful to describe the plasmon
satellites of silicon [\onlinecite{Reining,Guzzo2}], graphene [\onlinecite{Guzzo,Louie}],
and model systems [\onlinecite{Kas,Lischner2DEG}].
Recently, the application of cumulant expansion to the calculation of 
angle-resolved spectral function of simple semiconductors highlighted 
the dispersive character of plasmonic satellites [\onlinecite{Feliciano}].  
In particular, it was found that the simultaneous excitation of a plasmon and a hole leads to the 
emergence of 'plasmonic polaron bands', that are broadened replica of the 
valence bands blue-shifted by the plasmon energy. This observation was recently confirmed 
by theoretical and experimental work [\onlinecite{Caruso,Sky,Fadley}]. 
These recent developments call for a detailed analysis of cumulant expansion,
as well as its relation to the standard GWA.

The inequivalence of the results of GWA and cumulant approaches was demonstrated already in their earliest applications to the paradigmatic problem of photoemission spectra of core holes coupled to the linear electronic response of the medium [\onlinecite{BLundqvistGW}] for which cumulant expansion provides the exact solution [\onlinecite{Langreth}]. The existence of exact solution also enabled to estimate the error incurred by the application of a concrete form of GWA. In the studied core hole case [\onlinecite{BLundqvistGW}] the GWA solution was obtained from self-consistent calculation of the second order self-energy induced by the dynamically screened Coulomb interaction $W$ in the electron gas, and on the complete neglect of vertex corrections induced by $W$. This produced a compound structure in the core hole spectrum consisting of a narrow quasielastic peak at the 
blue-shifted core hole spectral threshold, and a much broader satellite or 'plasmaron' structure starting at plasmon frequency $\omega_{p}$ below the threshold and reaching the maximum further down at approximately $2\omega_{p}$ (cf. Fig. 2 in Ref. [\onlinecite{Langreth}]).

In contrast to the GWA, the second order cumulant expansion provides an exact multiexcitation solution for the spectrum of a core hole coupled to the boson-like dynamic component of $W$ [\onlinecite{Langreth}]. The obtained spectral structure exhibits the blue-shifted main quasiparticle peak that is followed by a series of satellites located at redshifted multiples of the plasmon excitation energy. All spectral peaks exhibit infra-red asymmetric broadening arising from the multiple excitation of low energy electron-hole (e-h) pairs present in the dielectric response of electron gas (cf. Fig. 4 in Ref. [\onlinecite{Langreth}]). Comparison of the cumulant and GWA solutions shows that the latter has the effect of forcing all the satellite peaks into a single one centered at the energy exceeding the plasmon excitation threshold by a factor $\geq 1.5$ and carrying their total weight.  This fundamental difference between the two types of spectra arises from the different classes of  processes the two approaches take into account. Cumulant expansion treats the processes described by boson-induced self-energy and vertex corrections on equivalent footing to all orders in the interaction and correlation between the successive scattering events. By contrast, the GWA solutions take into account only the self-energy corrections, either to lowest order(s) [\onlinecite{HedinPS}], or selfconsistently to all orders in the non-crossing boson interaction lines [\onlinecite{Engelsberg}], thereby leaving out all vertex corrections with crossing boson lines.  This means that in order to exploit the results of GWA calculations for construction of cumulants special care must be taken to select the appropriate GWA form which does not lead to overcounting of correlated higher order self-energy-like contributions in cumulant expansion. This requires unambigous identification of the levels of approximation at which GWA and cumulant expansion can be connected. Such a procedure is discussed in the forthcoming sections on the example of a single quasiparticle propagating in band states in which it is subject to the screened interaction $W$.  

While cumulant expansion has been discussed in previous works (cf. Refs. [\onlinecite{Mahan_book,Dunn,HedinPS,AlmbladhHedin,Bechstedt,4cumul,PSS2012}], given the renewed interest in this technique and its use in conjunction with the GWA[\onlinecite{AHK,Bechstedt,Reining,Louie,Feliciano,Fadley,Caruso,SGLD}] it seems appropriate and timely to systematically develop and discuss the key aspects of the GW + cumulant (GW+C) theory for interpretations of the results of both the stationary and time-resolved spectroscopies. This program is carried out in Secs. \ref{sec:G}-\ref{sec:L} and critically assessed in Sec. \ref{sec:GWlink}. For the benefit of GW practitioners who may want to directly employ the results of the presented formalisms without going into the details of its derivation we outline in Sec. \ref{sec:protocol} a protocol for a consistent combined use of the  {\it ab initio} GWA and cumulant expansion generated by the same dynamic interaction. 
Finally, as an illustrative example of the power of the developed approach we apply it in Sec. \ref{sec:Si}
to investigate the effects of electron interactions with the charge density fluctuations in the spectral function of silicon.

\section{Propagators of quasiparticles and bosonized response of the system}
\label{sec:G}

   A large class of many-body problems involving the interactions of quasiparticles (electrons or holes) with excitations in solids can be mapped onto the paradigmatic model of quasiparticle interactions with bosons. The justification for applying this model in the present studies of quasiparticles coupled to electronic excitations via the screened interaction $W$ is the bosonic character of its dynamic component [\onlinecite{Lazic_PRB,SGLD}].  Since our attention will be focused on excitations with energies ranging over the entire bandwidths we shall assume zero temperature of the system. Generalization to finite temperatures is a well established procedure.

 The generic Hamiltonian of the interacting particle-boson system reads
\bq
H=H^{0}_{\rm qp}+H^{0}_{bos}+\lambda V=H^{0}+\lambda V.
\label{eq:H}
\eq
We shall express the components of (\ref{eq:H}) in the second quantization form in terms of the fermion creation and annihilation operators $c_{\bf k}^{\dag}$ and $c_{\bf k}$ for the electron state with momentum ${\bf k}$, respectively, and the boson creation and annihilation operators $a_{{\bf q},j}^{\dag}$ and $a_{{\bf q},j}$ for the state of $j$-th mode with wavevector ${\bf q}$, respectively. Using these operators we have for the unperturbed Hamiltonian of the quasiparticles 
\bq
H^{0}_{\rm qp}=\sum_{\bf k}\epsilon^{0}_{\bf k}c_{\bf k}^{\dag}c_{\bf k},
\label{eq:Hqp}
\eq
where $\epsilon^{0}_{\bf k}$ in the unperturbed band state energy. The Hamiltonian of unperturbed bosons is  
\bq
H^{0}_{bos}=\sum_{{\bf q},j}\omega_{{\bf q},j}a_{{\bf q},j}^{\dag}a_{{\bf q},j},
\label{eq:Hbos}
\eq
where $\omega_{{\bf q},j}$ is the energy (dispersion) of the $j$-th mode. The particle-boson interaction is given by
\bq
\lambda V=\lambda \sum_{{\bf k,q},j}V_{\bf k-q,k}^{j}c_{\bf k-q}^{\dag}c_{\bf k}a_{{\bf q},j}^{\dag} + h.c.
\label{eq:V}
\eq

Spectral properties of quasiparticles (electrons and holes) whose dynamics is governed by the Hamiltonian (\ref{eq:H}) are most conveniently obtained from the diagonal component of causal (time ordered) quasiparticle propagator or Green's function $G_{\bf k}(t,t')$. At zero temperature this is defined by [\onlinecite{Mahan_book,Abrikosov,Zagoskin,Economou}]
\begin{widetext}
\barr
G_{\bf k}(t,t')&=&-i\langle 0|T[c_{\bf k}(t)c^{\dag}_{\bf k}(t')]|0\rangle
=
-i\theta(t-t')\langle 0|c_{\bf k}(t)c^{\dag}_{\bf k}(t')]0\rangle + i\theta(t'-t)\langle 0|c^{\dag}_{\bf k}(t')c_{\bf k}(t)|0\rangle
\label{eq:causalG}\\
&=& 
-i\theta(t-t')\int_{0}^{\infty}d\omega{\cal N}_{\bf k}^{>}(\omega)e^{-i(\omega+\mu)(t-t')}+i\theta(t'-t)\int_{0}^{\infty}d\omega{\cal N}_{\bf k}^{<}(\omega)e^{i(\omega-\mu)(t-t')},
\label{eq:causalGN}
\earr
\end{widetext}
where $|0\rangle$ denotes the ground state of the system, the operators $c^{\dag}_{\bf k}(t')$ and $c_{\bf k}(t)$ are expressed in the Heisenberg picture, and $T$ is the time ordering operator.
 The first term in (\ref{eq:causalG}) describes the probability amplitude that the electron created (injected) at time instant $t'$ in an initially unoccupied eigenstate $|{\bf k}\rangle$ of the unperturbed Hamiltonian $H^0$ of the system  is found in the same state at a later instant $t$. The second term gives the analogous probability for the hole created in an initially occupied state ${\bf k}$ at instant $t$ to be found in the same state at a later instant $t'$. Here ${\cal N}_{\bf k}^{>}(\omega)$ and ${\cal N}_{\bf k}^{<}(\omega)$ are the spectral densities of the particle and the hole, respectively, and $\mu$ is the chemical potential. 
In the absence of interactions of quasiparticles with the heatbath (\ref{eq:V}) the noninteracting quasiparticle Green's function reads
\barr
G_{\bf k}^{(0)}(t-t')= &-& i\theta(t-t')e^{-i(\epsilon_{\bf k}^{0}-i\eta)(t-t')}(1-n_{\bf k})\nonumber\\
&+&
i\theta(t'-t)e^{-i(\epsilon_{\bf k}^{0}+i\eta)(t-t')}n_{\bf k},
\label{eq:Gk0}
\earr
where $n_{\bf k}=\langle 0|c_{\bf k}^{\dag}c_{\bf k}|0\rangle$ is the occupation number of the one-particle ${\bf k}$-states in the ground state $|0\rangle$ and $\eta=0^{+}$.
This yields the Fourier transform (FT) of the unperturbed Green's function in the form
\bq
G_{\bf k}^{(0)}(\omega)=\frac{1-n_{\bf k}}{\omega-\epsilon_{\bf k}^0+i\eta}+ \frac{n_{\bf k}}{\omega-\epsilon_{\bf k}^0-i\eta}.
\label{eq:FTG0}
\eq

In the following we shall consider systems containing a {\it single} quasiparticle coupled to bosons. Since a single quasiparticle can not give rise to renormalization of the boson field [\onlinecite{Mahan_book}] the propagator $D_{\bf q}(t-t')$ of the $({\bf q},j)$-th boson mode in the system described by (\ref{eq:H}) should be equal to the unrenormalized one which at zero temperature takes a simple form [\onlinecite{Mahan_book}]
\barr
D_{{\bf q},j}^{(0)}(t-t')= &-& i\theta(t-t')e^{-i\omega_{{\bf q},j}(t-t')}\nonumber\\
&-&
 i\theta(t'-t)e^{i\omega_{{\bf q},j}(t-t')}.
\label{eq:Dq0}
\earr
However, in view of the forthcoming discussions of bosonic excitations of the electron gas we must assume a more general form
\barr
D_{\bf q}(t-t')=&-& i\theta(t-t')\int_{0}^{\infty}d\omega'{\cal D}_{\bf q}(\omega')e^{-i\omega'(t-t')}\nonumber\\
&-&
i\theta(t'-t)\int_{0}^{\infty}d\omega'{\cal D}_{\bf q}(\omega')e^{i\omega'(t-t')},
\label{eq:D}
\earr
where ${\cal D}_{\bf q}(\omega')$ is the mode density of bosonized heatbath excitations of wavevector ${\bf q}$ that lie in the energy interval $d\omega'$ around $\omega'$.  Thus, for quasiparticle interactions with the heatbath represented by the surrounding electron liquid, this is obtained from the imaginary part of the corresponding electron density-density response function $\chi_{\bf q}(\omega')$, viz. ${\cal D}_{\bf q}(\omega')=\frac{1}{\pi}|\mbox{Im}\chi_{\bf q}(\omega')|$. In this case we also have $V_{\bf k-q,k}^{j}=V_{\bf k-q,k}$. The thus formulated model describes equally well the propagation of quasiparticles in the bulk and in the quasi-two dimensional (Q2D) surface bands where they couple also to surface localized excitations (surface plasmons, surface phonons, etc.) [\onlinecite{PSS1984,PSS2012}].

Expressions (\ref{eq:FTG0}) and (\ref{eq:D}) together with the interaction matrix elements $V_{\bf k-q,k}$ represent the point of departure for calculations of $G_{\bf k}(t-t')$ renormalized by the interactions with bosonized electronic excitations in the system. Its FT yields the ${\bf k}$-resolved quasiparticle spectrum 
\bq
{\cal N}_{\bf k}(\omega)= \frac{1}{\pi} \left|\mbox{Im} G_{\bf k}(\omega)\right|={\cal N}_{\bf k}^{>}(\omega)\theta(\omega -\mu)+{\cal N}_{\bf k}^{<}(\omega)\theta(\mu-\omega).
 \label{eq:N}
\eq
The knowledge of (\ref{eq:N}) is needed in the determination of the various experimentally observable properties of the coupled quasiparticle-boson system. A characteristic example are photoemission yields from occupied electronic states of atomic, molecular and condensed matter systems which are directly related to the spectra of photoexcited holes [\onlinecite{Mahan_book,Langreth,Doniach,ChangLangreth,AlmbladhHedin,PSS1984,Ashcroft,Caroli,Inglesfield}].

Several methods of renormalization of (\ref{eq:Gk0}) by the interaction with bosons (\ref{eq:D}) to yield (\ref{eq:causalGN}) have been followed in the literature. The standard  procedure is based on the expansion of $G_{\bf k}$ in Dyson series involving the proper self-energy corrections $\Sigma_{\bf k}$ in ascending powers of the interaction [\onlinecite{Mahan_book,Abrikosov,Zagoskin,Economou}]. Here one of the most popular approximation schemes is the GW approximation in which the quasiparticle self-energy beyond the first order is represented by Feynman diagrams involving only noncrossing boson interaction lines (\ref{eq:D}). Depending on the complexity of the problem and the level of approximation [\onlinecite{MahanGW}], the effective number of interaction lines may range from one [\onlinecite{Hedin,BLundqvistGW,HedinPS}] to infinity in the self-consistent (SC) calculations[\onlinecite{Hedin,Engelsberg,Takada}]. The GWA results are considered to reliably reproduce the spectrum (\ref{eq:N}) near the quasiparticle excitation threshold, whereas they have been shown to be inadequate in the description of inelastic wings or satellite region [\onlinecite{Langreth,AHK}]. To remedy this deficiency of GWA another powerful method based on cumulant expansion [\onlinecite{Kubo,MHRT}] has been considered. The major rationale for this approach lies in the fact that cumulant expansion provides exact solutions in the limiting cases of (\ref{eq:H}) [\onlinecite{Langreth}], or a rapidly converging closed form solution in the general case [\onlinecite{AlmbladhHedin,Dunn,4cumul,PSS2012}].  Advantageous features of both approaches have been combined by a number of authors who have employed the GWA selfenergies as an input for cumulant representation of quasiparticle propagators so as to achieve a better description of satellite structure in the spectra [\onlinecite{AlmbladhHedin,AHK,Bechstedt,Lazic_PRL,Lazic_PRB,Reining,PSS2012,Louie,Feliciano,Fadley,Caruso,SGLD}]. However,  an explicit identification of the approximations underlying the combined use of GWA and cumulant expansion is still missing. In the following sections we make such a connection clear by using the model Hamiltonian (\ref{eq:H}).

The identification of a contact or "seam" between the GWA and cumulant expansion in a joint GW+C approach can be most clearly identified by resorting to a system for which a closed form solution for $G_{\bf k}(t,t')$ is available. Following this argument we analyse in the following the propagation of a {\it single} quasiparticle whose dynamics is governed by the Hamiltonian (\ref{eq:H}) [\onlinecite{PSS2012}]. In effect, this means that the considered electron (hole) is injected into an empty (occupied) band or into a state sufficiently above (below) the Fermi level.  This simplification introduces two important consequences in  the propagators (\ref{eq:causalG}) and (\ref{eq:FTG0}) describing the motion of quasiparticles: \\
(i) The quasiparticle propagates only in one time direction, i.e. only one term survives in (\ref{eq:causalG}), either the first one describing particles and proportional to $\theta(t-t')$ (implying the survival of only the first term in (\ref{eq:FTG0}) with $n_{\bf k}=0$), or the second one for holes and proportional to $\theta(t'-t)$  (implying analogously the survival of only the second term in (\ref{eq:FTG0}) with $n_{\bf k}=1$). 
\\ (ii) As a consequence of (i) all the quasiparticle-boson interaction vertices are restricted to the interval of quasiparticle propagation $(t,t')$. \\ Subject to these conditions cumulant expansion provides a tractable and asymptotically exact closed form solution for the quasiparticle propagators (see Sec. \ref{sec:propagators_cumul} below) which then allows equally tractable comparison with the results of GWA approach.

\section{Quasiparticle self-energy in the $G_{0}W_{0}$ approximation}
\label{sec:GW}

A convenient representation of the general solution for Green's function (\ref{eq:causalG}) in the case of homogeneous and conservative systems is usually formulated for its Fourier transform $G_{\bf k}(\omega)$.  The integral Dyson equation for diagonal Green's function in the four-coordinate space transforms into algebraic equation in the reciprocal $({\bf k},\omega)$-space where it can be represented in  the form [\onlinecite{Mahan_book,Abrikosov,Zagoskin,Economou}]
\bq
G_{\bf k}(\omega)=G_{\bf k}^{(0)}(\omega)+G_{\bf k}^{(0)}(\omega)\Sigma_{\bf k}(\omega)G_{\bf k}(\omega).
\label{eq:GSigma}
\eq
Here the proper self-energy $\Sigma_{\bf k}(\omega)$ is given by the sum of all irreducible components involving the powers of interactions that perturb the initial noninteracting Hamiltonian $H^0$.

In the first application of GWA the lowest term of proper self-energy of the interacting electron gas was written "on-the-energy-shell" $\omega=\epsilon_{\bf k}^{0}$ in the form [\onlinecite{QF1958}]
%
\barr
\Sigma_{\bf k}^{G_{0}W_{0}}(\epsilon_{\bf k}^{0})&=&\int \frac{d^3 q}{(2\pi)^3}\int \frac{d\omega'}{2\pi}\left.G^{(0)}_{\bf k-q}(\omega-\omega')W_{\bf q}^{(0)}(\omega'))\right|_{\omega=\epsilon_{\bf k}^0}\nonumber\\
&=&
\int \frac{d^3 q}{(2\pi)^3}\int \frac{d\omega'}{2\pi}\frac{v_{\bf q}}{\varepsilon_{\bf q}(\omega')}\left.\frac{1}{\omega-\omega'-\epsilon_{\bf k-q}^0\pm i\eta}\right|_{\omega=\epsilon_{\bf k}^0},
\label{eq:QFSigma}
\earr
%
and subsequently elaborated "off-the-energy-shell" [\onlinecite{Hedin}] and termed $G_{0}W_{0}$ self-energy [\onlinecite{MahanGW}]. Here $G^{(0)}$ denotes the unperturbed propagator of the electron or hole with $\pm i\eta$ in the denominator, respectively, and 
\bq
W_{\bf q}^{(0)}(\omega')=v_{\bf q}/\varepsilon_{\bf q}(\omega')
\label{eq:W}
\eq
stands for the matrix element of the bare Coulomb interaction $v_{\bf k-q,k}=v_{\bf q}$ linearly screened by the dielectric function of the electron gas $\varepsilon_{\bf q}(\omega')$.
\begin{figure}[tb]
\rotatebox{0}{ \epsfxsize=8cm \epsffile{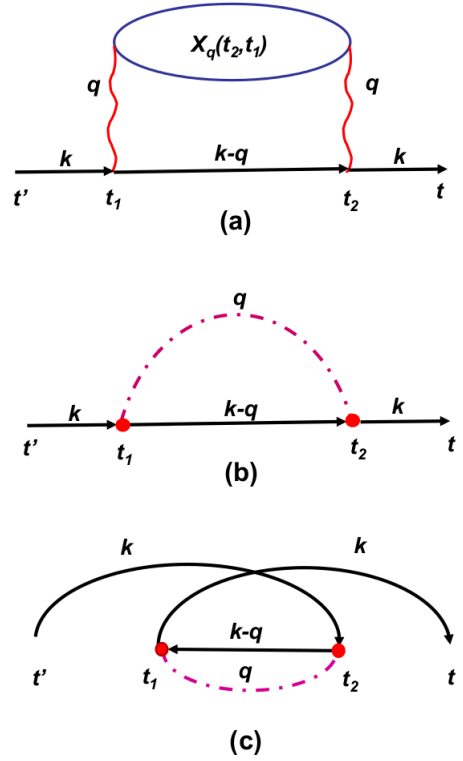} } 
\caption{(Color online)(a) Diagram for the second order contribution in the expansion of Green's function (\ref{eq:causalG}) of an electron injected with the initial wavevector ${\bf k}$ into an empty band. Full lines denote the
unperturbed propagators, wavy lines the unscreened Coulomb interaction,
and the bubble denotes the electronic response function $\chi_{\bf q}(t_{2},t_{1})$ of
the system.  (b) Equivalent diagram in the quasiparticle-boson
 model outlined in Sec. \ref{sec:G}. Full circles denote the interaction matrix element $v_{\bf q}$ from Eq. (\ref{eq:Wchi}) and dash-dotted line denotes the boson propagator $D_{\bf q}(t_{2}-t_{1})$ in the intermediate state interval $(t_{1},t_{2})$. (c) Example of second order diagram involving two electrons above and one hole below the Fermi level in the same interval. This term can not be generated within the single particle limit of the Hamiltonian (\ref{eq:H}).}
\label{Sigma_k}
\end{figure}

Using (\ref{eq:QFSigma}) as a point of departure we can invoke the definition of dielectric function in the linear response formalism  and write (\ref{eq:W}) in the form
\bq
W_{\bf q}^{(0)}(\omega')=v_{\bf q}+v_{\bf q}\chi_{\bf q}(\omega')v_{\bf q},
\label{eq:Wchi}
\eq
where $\chi_{\bf q}(\omega')$ is the electronic density-density response function calculated in the standard RPA or one of its generalized versions [\onlinecite{SGLD}]. Now, it is easy to verify that the function $\chi_{\bf q}(\omega')$ calculated within the linear response formalism has the properties of a boson propagator (\ref{eq:D}) with the spectrum of excitations given by ${\cal D}_{\bf q}(\omega')=(1/\pi)|\mbox{Im}\chi_{\bf q}(\omega')|$ [\onlinecite{Lazic_PRL}]. This enables to represent the off-the-energy-shell form of expression (\ref{eq:QFSigma}) as
\bq
\Sigma_{\bf k}^{G_{0}W_{0}}(\omega)=\Sigma^{(1)}_{\bf k}+\Sigma^{(2)}_{\bf k}(\omega).
\label{eq:SigmaGW}
\eq
The first component on the RHS of (\ref{eq:SigmaGW}) reads
\bq
\Sigma^{(1)}_{\bf k}=\int \frac{d^4 q}{(2\pi)^{3}} v_{\bf q}\int \frac{d\omega'}{2\pi}G^{(0)}_{\bf k-q}(\omega-\omega')
\label{eq:Sigma1}
\eq
and represents the $\omega$-independent Fock-like correction equal to the first order exchange self-energy [\onlinecite{Mahan_book}]. This static contribution can be associated with other static contributions that renormalize $\epsilon_{\bf k}^{0}$, but it will be of no further interest in the context of dynamical particle-boson interactions studied within the cumulant approach. On the other hand, the second component (upper and lower signs stand for electrons and holes, respectively)
\bq
\Sigma^{(2)\stackrel{>}{\scriptscriptstyle <}}_{\bf k}(\omega)=\int \frac{d^3 q}{(2\pi)^3}\int d\omega' \frac{v_{\bf q}^{2}{\cal D}_{\bf q}(\omega')}{\omega-\epsilon_{\bf k-q}^{0} \mp \omega'\pm i\eta},
\label{eq:Sigma2k}
\eq
is dynamic as it gives $\omega$-dependent contribution to (\ref{eq:SigmaGW}) shown diagrammatically in the time domain as a self-energy insertion in Fig. \ref{Sigma_k}a. This has also been referred to as the correlation part or polarization contribution in Refs. [\onlinecite{AryasetiawanGunnarsson}] and [\onlinecite{Hedin1999}], respectively.
 Therefore, the term (\ref{eq:Sigma2k}) can be considered as describing the quasiparticle (electron or hole) interaction with the field of bosonized excitations of the surrounding electron gas. This passage is illustrated in Fig. \ref{Sigma_k}a-c. 

Expressions (\ref{eq:QFSigma})-(\ref{eq:Sigma2k}) will enable us to establish a rigorous correspondence between the partial solution of the problem of dynamic interactions of a quasiparticle obtained within the GWA and cumulant approach.  We note that it is also possible to extend the $G_{0}W_{0}$ approximation by dressing $G^{(0)}$ through the self-energy insertions induced by $W^{(0)}$, and solve self-consistently the equation for the thus defined $G$ [\onlinecite{MahanGW,Engelsberg}]. First-principles calculations using the self-consistent $GW$ formalism were reported in Refs. [\onlinecite{Holm,Stan,Thygesen,Caruso-1,Caruso-2}].
However, we shall demonstrate in the next sections that the results of such higher order renormalizations can not be mapped onto the standard cumulant expansion in ascending powers of the coupling constant  $\lambda$ [\onlinecite{Kubo,Mahan_book}]. Hence, they should not be pursued in the combined GW+C treatment of quasiparticle propagators.

\section{Single quasiparticle propagators from cumulant expansion }
\label{sec:propagators_cumul}

\subsection{Cumulant expansion for a coupled particle-boson system}
\label{sec:particle-boson}

The dynamics of a quasiparticle injected into a state in an empty band state of the system is conveniently extracted from the quasiparticle propagator (\ref{eq:G}). Here we follow the notation and conventions of Refs. [\onlinecite{MHRT}] and [\onlinecite{Mahan_book}], and for the sake of easier visualization derive first the cumulants for electron propagators in the time domain. The cumulants for the propagators describing hole dynamics in occupied bands are derived by analogy in subsection \ref{sec:C2properties}.

We start from the Hamiltonian (\ref{eq:H}) and assume  that the eigenstates $\{|{\bf k}\rangle\}$ of  $H_0$ which define the initial one-electron band structure of the system are known, i.e. that this part of the problem has been solved first.
 In the single particle limit the ground state average in (\ref{eq:causalG}) can be conveniently calculated by using two standard hints. First, on noticing that before the injection of the particle into a band state described by the quantum number ${\bf k}$  one has $H|0\rangle=H^0|0\rangle=E^0|0\rangle$. This is in contrast to many-particle systems in which the initial ground state $|0\rangle$ is also renormalized by the interaction $\lambda V$. Second, the evolution operator governing the time dependence of the operators $c^{\dag}_{\bf k}(t)=e^{iHt}c^{\dag}_{\bf k}e^{-iHt}$ and $c^{\dag}_{\bf k}(t)=e^{iHt}c_{\bf k}e^{-iHt}$ is represented in the form
\bq
e^{-iHt}=e^{-i(H^0+\lambda V)t}=U(t)=e^{-iH^0 t}U_{I}(t).
\label{eq:U}
\eq
Here $U_{I}(t)=e^{iH^0 t}e^{-iH t}$ is the evolution operator in the interaction picture generated by the perturbation $\lambda V$ and satisfying the integral equation
\bq
U_{I}(t)=1-i\int_{0}^{t} dt'V_{I}(t')U_{I}(t'),
\label{eq:U_I}
\eq
where $V_{I}(t')=e^{iH^0 t'}\lambda Ve^{-iH^0 t'}$.  Now, since the unperturbed  single particle Green's function in the absence of the interaction $\lambda V$ reads
\bq
G_{\bf k}^{(0)}(t-t')=-i\theta(t-t')e^{-i(\epsilon^{0}_{\bf k}-i\eta)(t-t')}.
\label{eq:G0}
\eq
we can write the full single particle Green's function (\ref{eq:causalG}) 
\bq
G_{\bf k}(t-t')=G_{\bf k}^{(0)}(t-t')\langle {\bf k}|U_{I}(t-t')|{\bf k}\rangle,
\label{eq:G}
\eq
where in shorthand notation $|{\bf k}\rangle=c_{\bf k}^{\dag}|0\rangle$. Thus the calculation of the single particle propagator deriving from (\ref{eq:causalG}) reduces to the evaluation of the diagonal matrix elements of $U_I(t)$ in the unperturbed basis.  Thereby a perturbative solution to (\ref{eq:causalG}) involves the unperturbed propagators   (\ref{eq:D}) and (\ref{eq:G0}) and the matrix elements of $\lambda V$.

The evolution operator $U_I(t)$ can always be expressed in an exponential form [\onlinecite{Gross}] and the averages of such generalized exponential operators are most conveniently calculated by using cumulant expansion in powers of the coupling constant $\lambda$ of the perturbation that generates $U_I(t)$ [\onlinecite{Kubo}]. This gives [\onlinecite{Gross,Doniach_book}]
\bq
\langle {\bf k}|U_{I}(t-t')|{\bf k}\rangle=e^{C_{\bf k}(t-t')}
\label{eq:UI}
\eq
and hence
\bq
G_{\bf k}(t-t')=G_{\bf k}^{(0)}(t-t')e^{C_{\bf k}(t-t')}.
\label{eq:Gcumul}
\eq
In the present problem defined by the Hamiltonian (\ref{eq:H}) the cumulant function
\bq
C_{\bf k}(t-t')=\sum_{l=2}^{\infty} C_{\bf k}^{(l)}(t-t')
\label{eq:Cl}
\eq
is an infinite ascending series of cumulants in even powers of $\lambda$, viz. $C_{\bf k}^{(l)}(t-t')\propto \lambda^{l}$. They can be evaluated by directly employing the cumulant averaging of time-ordered products of operators $V_{I}(t)$ [\onlinecite{Kubo,Dunn,4cumul}], or by equating the $\lambda^{l}$-th order terms of the Dyson expansion for the Green's function on the LHS with the same powers of expanded exponential on the RHS of Eq. (\ref{eq:Gcumul}) [\onlinecite{Mahan_book,Meden}].  
The series (\ref{eq:Cl}) can converge fast even for strong coupling provided the correlations between successive quasiparticle-boson scattering events in higher order cumulants are small [\onlinecite{correlations,vanHaeringen}], e.g. under the conditions of small relative transfers of energy and momentum from or to the quasiparticle during its propagation [\onlinecite{Dunn,4cumul,PSS2012}]. This is a consequence of a general theorem [\onlinecite{Kubo}] which states that higher order cumulants are expressed as differences between correlated and uncorrelated averages of the time-ordered products of operators $V_{I}(t)$. Thus, in the limit of infinite quasiparticle mass, which is realized for holes created in deep core levels of solids or adsorbates, already the second order cumulant provides an exact result to (\ref{eq:Gcumul}) and (\ref{eq:Cl}) [\onlinecite{Langreth,AlmbladhHedin,PSS1984}]. This is in stark contrast to the standard Dyson expansion of Green's function (or the corresponding self-energy) where such additional small expansion parameter does not exist. This has a very important practical implication in that the approximate Green's function (\ref{eq:Gcumul}) calculated by including only a small number of low order cumulants gives a very accurate result, whereas the Dyson expansion of similar accuracy usually requires an infinite number of terms.

From the general properties of the evolution operator (\ref{eq:U_I}) generated by the Hamiltonian (\ref{eq:H}) one derives the following relations [\onlinecite{PSS2012}] satisfied by the cumulants (\ref{eq:Cl}) 
\barr
C_{\bf k}(0)=0,
\label{eq:unitarity}\\
\dot{C}_{\bf k}(0)=0,
\label{eq:zerowork}
\earr
where dot denotes the time derivative. The properties (\ref{eq:unitarity}) and (\ref{eq:zerowork}) apply to the whole series (\ref{eq:Cl}) as well as separatelly  to its constituents. Expression (\ref{eq:unitarity}) reflects the conservation of the norm of the spectrum ${\cal N}_{\bf k}(\omega)$ (unitarity). Relation (\ref{eq:zerowork}) is a consequence of the property $C_{\bf k}^{(l\geq 2)}(t\rightarrow 0)\propto t^{l}$ arising from the generating interaction (\ref{eq:V}), and reflects the conservation of the first moment of the spectrum ${\cal N}_{\bf k}(\omega)$ (zero work sum rule [\onlinecite{LangrethNobel,PSS2012}]).

For the quasiparticle-boson model interaction (\ref{eq:V}) the second order cumulant $C_{\bf k}^{(2)}(t)$ can be readily calculated [\onlinecite{PSS2012}] and for further convenience and reference its explicit derivation will be presented in subsection \ref{sec:2cumul}. Folllowing the procedures outlined in Ref. [\onlinecite{Kubo}] the contributions from higher, fourth order cumulants were calculated in Refs. [\onlinecite{Dunn}] and [\onlinecite{4cumul}] for nondispersive and dispersive bosons, respectively, and  for all reasonable values of the coupling constants they have in both cases been found negligible relative to the ones from second order cumulants.  This property has been widely assumed as a general feature in all later modellings of single quasiparticle interactions with bosonic type of excitations based on cumulant expansion [\onlinecite{Doniach,MHRT,AHK,Bechstedt,Lazic_PRL,Reining,PSS2012,Louie,Feliciano,Fadley,Caruso}].  This approach has also proved extremely useful in the descriptions  of ultrafast phenomena that can be revealed by time resolved electron spectroscopies [\onlinecite{Lazic_PRL,Lazic_PRB,PSS2012,NP2014,SGLD}].

At this point a note is in order on the use of unperturbed quasiparticle Hamiltonians $H_{\rm qp}$ which are diagonal on the Kohn-Sham (KS) one particle state basis and thus embody the electronic exchange-correlation (xc) effects in the corresponding KS one particle energies. In order to avoid in this case the overcounting of xc-effects in Eqs. (\ref{eq:Gcumul}) and (\ref{eq:Cl}) one should introduce the first order cumulant $C_{\bf k}^{(1)}(t-t')$ which subtracts these effects from the renormalized energies in $G_{\bf k}$. This is analogous to the treatment of static initial state interactions described in Fig. 8b of Ref. [\onlinecite{Langreth}] and in Sec. II.B. of Ref. [\onlinecite{Lazic_PRB}]. In the case of KS states obtained within the LDA scheme this is achieved by extending  (\ref{eq:Cl}) with the term
\bq
C_{\bf k}^{(1)}(t-t')=-i\left(-v_{\bf k}^{xc}(t-t')\right),
\label{eq:C1}
\eq
where $v_{\bf k}^{xc}$ is the diagonal matrix element of the local KS LDA exchange-correlation potential in the state ${\bf k}$. This gives a stationary contribution to the quasiparticle energy that can be associated with $\epsilon_{\bf k}^{0}$ in $G_{\bf k}^{(0)}$, thereby allowing further treatment of the dynamic features of cumulant expansion without limitations imposed on the unperturbed basis. Note in passing that the property (\ref{eq:zerowork}) does not apply to the {\it ad hoc} added cumulant (\ref{eq:C1}).

\subsection{Second order cumulant}
\label{sec:2cumul}

The Feynman diagram corresponding to the second order process in $\lambda V$ that is common to cumulant and Dyson expansion of Green's function (\ref{eq:causalG}) governed by (\ref{eq:H}) is shown in Fig. 1b. It involves the unperturbed propagators (\ref{eq:D}) and (\ref{eq:G0}), and the matrix elements $V_{\bf k-q,k}$ of the interaction $V$ in the basis of quasiparticle and boson eigenfunctions diagonalizing $H^{0}$. For interactions (\ref{eq:V}) constrained to the interval $(t',t)$ this gives the first nonvanishing correction to the unperturbed single electron Green's function to order $\lambda^2$:
\begin{widetext}
\barr
iG^{(2)}_{\bf k}(t-t')&=&
\int_{t'}^{t}d t_2 \int_{t'}^{t} d t_1 iG^{(0)}_{\bf k}(t-t_2)\left[\sum_{\bf q}(-i\lambda V_{\bf k-q,k})^{2}iD_{\bf q}(t_2-t_1)iG^{(0)}_{\bf k-q}(t_2-t_1)\right]iG^{(0)}_{\bf k}(t_1-t')
\label{eq:GSigma2}\\
&=&
iG^{(0)}_{\bf k}(t-t')\left[\sum_{\bf q}(-i\lambda V_{\bf k-q,k})^{2}\int d\omega' {\cal D}_{\bf q}(\omega')\frac{1-e^{-i(\epsilon^{0}_{\bf k-q}+\omega'-\epsilon^{0}_{\bf k})(t-t')}-i(\epsilon^{0}_{\bf k-q}+\omega'-\epsilon^{0}_{\bf k})(t-t') }{(\epsilon^{0}_{\bf k-q}+\omega'-\epsilon^{0}_{\bf k})^2}\right]
\label{eq:GC2def}\\
&=&
iG^{(0)}_{\bf k}(t-t')C_{\bf k}^{(2)}(t-t').
\label{eq:GC2}
\earr
\end{widetext}
To obtain (\ref{eq:GC2def}) and (\ref{eq:GC2}) we have exploited the single particle-imposed time ordering $t>t_{2}>t_{1}>t'$.
The thus obtained $G^{(2)}_{\bf k}(t-t')$ serves as a definition of the second order self-energy given in the time domain through expression in the square bracket on the RHS of (\ref{eq:GSigma2}),  and of the second order cumulant generated by the same interaction and given by expression in the square bracket in (\ref{eq:GC2def}). Respectively, they are given by the expression 
\bq
-i\Sigma_{\bf k}^{(2)}(t_2-t_1)=\sum_{\bf q}(-i\lambda V_{\bf k-q,k})^{2}iD_{\bf q}(t_2-t_1)iG^{(0)}_{\bf k-q}(t_2-t_1),
\label{eq:defSigma2}
\eq
which constitutes the diagram in Fig. \ref{Sigma_k}b and  whose FT yields (\ref{eq:Sigma2k}), and 
\begin{widetext}
\bq
C_{\bf k}^{(2)}(t-t')=-\sum_{\bf q}(\lambda V_{\bf k-q,k})^{2}\int d\omega' {\cal D}_{\bf q}(\omega')
\frac{1-e^{-i(\epsilon^{0}_{\bf k-q}+\omega'-\epsilon^{0}_{\bf k})(t-t')}-i(\epsilon^{0}_{\bf k-q}+\omega'-\epsilon^{0}_{\bf k})(t-t')}{(\epsilon^{0}_{\bf k-q}+\omega'-\epsilon^{0}_{\bf k})^2}.
\label{eq:C2}
\eq
\end{widetext}
The factorization appearing in the last line of (\ref{eq:GC2}) is a direct consequence of the simple exponential form of  unperturbed $G^{(0)}_{\bf k}(t)$ given by (\ref{eq:G0}) which allows propagation in one time direction only. Such a factorization can be established in  all orders of perturbation and enables the rearrangement of the Dyson series for a single particle propagator into a product of $iG^{(0)}_{\bf k}(t-t')$ and the terms that can be directly related to exponentiated sum of cumulants (\ref{eq:Cl}) [\onlinecite{Mahan_book,Meden}]. 
We stress that this property does not generally hold for causal, time ordered quasiparticle propagators; therefore time-ordered propagators cannot be represented in simple cumulant form.
This implies that the use of the cumulant expansion is fully justified only when considering the
quasiparticle Green's functions propagating in one time direction only [\onlinecite{ND,Langreth,Doniach,Dunn,AlmbladhHedin,4cumul,PSS2012,Kas}].

Another important property of the thus derived $G^{(2)}_{\bf k}(t-t')$ is the absence of the factor $i\eta$ from explicit expression for $C_{\bf k}^{(2)}(t-t')$ and its presence only in the factorized term $G^{(0)}_{\bf k}(t-t')$. This is due to cancellations of exponents in the products of factors $e^{-\eta t_j}e^{+\eta t_j}$ arising from two successive $G^{(0)}_{\bf k}$s in each $j$-th interaction vertex. Note that for this to occur $\eta$ need not be infinitesimal but only constant. Therefore, in order to study the temporal properties of $C_{\bf k}^{(2)}(t-t')$ it would be {\it inappropriate} to introduce at the outset the factor $i\eta$ into the denominator on the RHS of (\ref{eq:C2}), the more so as the whole expression is nonsingular for $(\epsilon^{0}_{\bf k-q}+\omega'-\epsilon^{0}_{\bf k})\rightarrow 0$. These subtle properties of $C_{\bf k}^{(2)}(t-t')$ are further elaborated below.

\subsection{Integral representation of $C_{\bf k}(t-t')$}
\label{sec:C2properties}

Expression (\ref{eq:C2}) can be brought to a more compact form by making use of definition (\ref{eq:D}) and by introducing the interaction weighted joint density of excitations $\rho_{\bf k}^{(2)>}(\nu)$ for electron propagation  with $\epsilon_{\bf k}^{0}$ and $\epsilon_{\bf k-q}^{0}$ above the Fermi level [\onlinecite{MHRT,PSS2012}]. Introducing the matrix elements $\lambda V_{\bf k-q,k}=v_{\bf k-q,k}$ we obtain by using (\ref{eq:D}) and (\ref{eq:Sigma2k})
\begin{widetext}
\bq
\rho_{\bf k}^{(2)>}(\nu)=\sum_{\bf q}v_{\bf k-q,k}^2(1-n_{\bf k})(1-n_{\bf k-q})\int_{0}^{\infty} d\omega'{\cal D}_{\bf q}(\omega')\delta(\nu-(\omega'+\epsilon^{0}_{\bf k-q}-\epsilon^{0}_{\bf k})) =
-\frac{1}{\pi}\mbox{Im}\Sigma_{\bf k}^{(2)>}(\epsilon_{\bf k}^{0}+\nu),
\label{eq:rho2>}
\eq
\end{widetext}
The thus defined $\rho_{\bf k}^{(2)>}(\nu)$ is bounded from below at $\nu=-(\epsilon^{0}_{\bf k}-E_{\rm low})\leq 0$ when the particle is scattered to the lowest unoccupied state of energy $E_{\rm low}$ which is the supremum of the band bottom energy $\epsilon^{0}_{\bf k-q}=0$ and the Fermi energy $\epsilon_{F}=\mu$. The other limit $\rho_{\bf k}^{(2)>}(\nu\rightarrow\infty)$ is dictated by the asymptotic behavior of $\mbox{Im}\chi_{\bf q}(\nu)$ [\onlinecite{Giuliani}].  

In the case of single hole propagation (symbol $<$) the corresponding joint density of excitations takes the form
\begin{widetext}
\bq
\rho_{\bf k}^{(2)<}(\nu)=\sum_{\bf q}v_{\bf k-q,k}^2n_{\bf k}n_{\bf k-q}
\int_{0}^{\infty} d\omega'{\cal D}_{\bf q}(\omega')\delta(\nu-(\omega'-\epsilon^{0}_{\bf k-q}+\epsilon^{0}_{\bf k}))=
\frac{1}{\pi}\mbox{Im}\Sigma_{\bf k}^{(2)<}(\epsilon_{\bf k}^{0}-\nu),
\label{eq:rho2<}
\eq
\end{widetext}
and is lower bounded at $\nu=-(E_{\rm up}-\epsilon_{\bf k}^{0})\leq 0$ where $E_{\rm up}$ is the infimum of the upper occupied band edge and $\epsilon_{F}$.

Definitions (\ref{eq:rho2>}) and (\ref{eq:rho2<}) enable us to write the second order cumulant for electron and hole propagation in a compact form
\barr
C_{\bf k}^{(2)\stackrel{>}{\scriptscriptstyle <}}(t)&=&-\int d\nu \rho_{\bf k}^{(2)\stackrel{>}{\scriptscriptstyle <}}(\nu)\frac{1-e^{\mp i\nu t} \mp i\nu t}{\nu^2}\nonumber\\
&=&
-\int d\nu \left( \mp \frac{1}{\pi} \Sigma_{\bf k}^{(2)\stackrel{>}{\scriptscriptstyle <}}(\epsilon_{\bf k}^{0}\pm \nu) \right)\frac{1-e^{\mp i\nu t} \mp i\nu t}{\nu^2}.
\label{eq:C_rho2}
\earr
Here it should be observed that $\rho_{\bf k}^{(2)\stackrel{>}{\scriptscriptstyle <}}(\nu)$ are not on-the-energy-shell quantities so that the energy conservation contained in $C_{\bf k}^{(2)\stackrel{>}{\scriptscriptstyle <}}(t)$ is determined solely by the properties of long time limit $t\rightarrow \infty$ of expression (\ref{eq:C_rho2}).

Although the integral representation (\ref{eq:C_rho2}) of the cumulant function  has been here established for the second order term in (\ref{eq:Cl}), it can be shown (see Sec. A of Supplemental material [\onlinecite{Suppl}]) that the whole cumulant series (\ref{eq:Cl}) can be formally obtained through the mapping $\rho(\nu)\leftrightarrow C(t)$ using the same integral transform as in (\ref{eq:C_rho2}) [\onlinecite{PSS2012}], viz.
\bq
C_{\bf k}^{\stackrel{>}{\scriptscriptstyle <}}(t)=-\int d\nu \rho_{\bf k}^{\stackrel{>}{\scriptscriptstyle <}}(\nu)\frac{1-e^{\mp i\nu t} \mp i\nu t}{\nu^2}.
\label{eq:C_rho}
\eq
In both cases (hereafter in the present subsection we drop the superscripts $>$ and $<$)
\bq
\rho_{\bf k}(\nu)=\sum_{l=2}^{\infty}\rho_{\bf k}^{(l)}(\nu),
\label{eq:rho}
\eq
 with $\rho_{\bf k}^{(l)}(\nu)\propto\lambda^{l}$, and where $\rho_{\bf k}^{(2)}(\nu)$ is the lowest term of the series  from which (\ref{eq:C1}) is excluded. Owing to the bijection properties of the integral transform (\ref{eq:C_rho}) here each $\rho_{\bf k}^{(l)}(\nu)$ can be uniquely obtained from the inverse transform of the corresponding $C_{\bf k}^{(l)}(t)$ [\onlinecite{PSS2012}] defined by
\bq
\rho_{\bf k}^{(l)}(\nu)=-\frac{1}{2\pi}\int_{-\infty}^{\infty}\ddot{C}_{\bf k}^{(l)}(t)e^{i\nu t}dt, \hspace{5mm} l>1,
\label{eq:ddotC}
\eq
where double dot denotes second order time derivative. 
Expressions (\ref{eq:C_rho}) and (\ref{eq:rho}) also enable a compact  semi-diagonal representation of the initial Hamiltonian (\ref{eq:H}) in terms of particle interaction with {\it drag bosons} formulated  in Sec. B of Supplemental material [\onlinecite{Suppl}]. This representation facilitates direct constructions of the diagonal single quasiparticle propagators (\ref{eq:causalG}) and cumulant series (\ref{eq:Cl}).

Despite the fact that (\ref{eq:ddotC}) now extends the applicability of (\ref{eq:C_rho2}) to the entire series (\ref{eq:Cl}), it again poses the question of tedious derivations of higher order cumulants and the ensuing $\rho_{\bf k}^{(l)}(\nu)$ by the standard cumulant expansion of the evolution operator [\onlinecite{Kubo,4cumul}]. Therefore, the importance of the general integral representation defined by (\ref{eq:C_rho}), (\ref{eq:rho}) and (\ref{eq:ddotC}), and proven in Sec. A of Supplemental material [\onlinecite{Suppl}], is in the domain of the existential theorem  for representation of the cumulant series generated by the Hamiltonian (\ref{eq:H}), and the discussion of its global properties in the various temporal intervals of physical relevance (see Sec. \ref{sec:L}). However, as will be pointed out in Sec. \ref{sec:validityC2}, all higher order terms give rise to negligible corrections in the case of systems that dynamicswise can be mapped to particle-boson model Hamiltonians with standard coupling strengths [\onlinecite{Dunn,4cumul}].

\subsection{Range of validity and limitations of the second order cumulant expansion}
\label{sec:validityC2}

Cumulant expansion is of great practical importance for the class of problems or model systems for which the low order cumulants provide either the full or dominant contribution to the series (\ref{eq:Cl}).  A typical example of the former case is the core hole problem described by (\ref{eq:H}) with the single core level Hamiltonian $H^{0}_{c}=\epsilon^{0}_{c}c^{\dag}_{c}c_{c}$ and the factorized coupling of the hole density to the boson field, viz. $\lambda V\propto -c_{c}c^{\dag}_{c}$ (see Refs. [\onlinecite{Langreth,PSS1984}]). Here the sum of the first and second order cumulants provides an exact solution due to the absence of any correlations between the successive boson emission and absorption events because the source of perturbation on the boson field, i.e. the localized core hole, is dynamically structureless.   Its only degree of freedom is its energy $\epsilon_{c}$ which, likewise the matrix elements $v_{\bf k-q,k}\rightarrow v_{\bf q}$, is not affected by the quasiparticle recoil in the boson excitation processes. This gives rise to cancellation of all higher order cumulants which measure correlations among the interaction vertices [\onlinecite{Kubo}] and to reduction of cumulant series (\ref{eq:Cl}) to the second order term (\ref{eq:C_rho2}).  This feature has been successfully employed in the interpretations of core level lineshapes in the bulk [\onlinecite{Langreth,AlmbladhHedin}] and at surfaces [\onlinecite{PSS1984}].

The above discussion gives a clue as to the validity of approximate treatment of the problem posed by quasiparticle-boson Hamiltonian (\ref{eq:H}). Inspection of the expressions for the fourth order cumulants [\onlinecite{Dunn,4cumul}] shows that in the presence of translational  invariance in the phase space which applies to the energy differences
\bq
\epsilon^{0}_{\bf k+q+p}-\epsilon^{0}_{\bf k+q}\leftrightarrow \epsilon^{0}_{\bf k+p}-\epsilon^{0}_{\bf k},
\label{eq:trans_epsilon}
\eq
and to the interaction matrix elements
\bq V_{\bf k+q+p,k+q}\leftrightarrow V_{\bf k+p,k},
\label{eq:transV}
 \eq
the fourth and higher order cumulants turn to zero. These invariances are characteristic of the absence of correlations
between successive boson emission and reabsorption events.  Hence, in the complete absence of such correlations,
as is the case with boson fields perturbed either by structureless or classical time
dependent potentials, the cumulant series (\ref{eq:Cl}) reduces to a single term given by
the second order cumulant (\ref{eq:C2}) in which $V_{\bf k-q,k}\rightarrow V_{\bf
q}$ and $\epsilon^{0}_{\bf k-q}-\epsilon^{0}_{\bf k}\rightarrow 0$. This exactly solvable limit of the particle-boson Hamiltonian is known as the forced oscillator model [\onlinecite{Mahan_book,MHRT,BreGum,Sunko}]. If the correlations are nonvanishing but small, then the second order cumulant expansion provides a good approximation for the calculation of quasiparticle amplitudes (\ref{eq:Gcumul}) [\onlinecite{Dunn,4cumul}].

The translational invariaces (\ref{eq:trans_epsilon}) and (\ref{eq:transV}) are expected to hold well in two extreme situations. The first is the case of very large quasiparticle mass, or equivalently very narrow quasiparticle band, which is the core hole limit elaborated above. The second is the case of very broad quasiparticle band and initial $\epsilon^{0}_{\bf k}$ far from the band edges (or from the Fermi level).  In this case the variations of $\epsilon^{0}_{\bf k}$ may be considered to be nearly linear, and the variation of the matrix elements of $\lambda V$ nearly zero for small variations of the quantum number $|{\bf k}|$, which altogether makes all $C^{(l>2)}_{\bf k}(t)$ very small. Therefore, the convergence of the cumulant series (\ref{eq:Cl}) is system specific as it depends on the trade off between the strength of the interaction (i.e. the magnitude of the coupling constant $\lambda$) and the correlations between successive scattering events.

Second order cumulant expansion becomes increasingly worse as the quasiparticle energy moves closer to the band edges where correlations between subsequent scattering events induced by $\lambda V$ may become large. In this regime the low order self-energy corrections may represent a better approximation to the quasiparticle propagator, particularly in relation to its asymptotic time behavior characterized by the so-called quasiparticle collapse [\onlinecite{Khalfin,PSS2012,Pastawski,impurity_scatt,Marion}]. Moreover, near the Fermi level the single particle approximation which separates electron from hole motion and leads to (\ref{eq:rho2>}) and (\ref{eq:rho2<}) becomes unrealistic and the processes like those shown in Fig. \ref{Sigma_k}c need be considered. In principle, there is no difficulty with their implementation because they also generate the same form of second order cumulant provided $t_{1}$ and $t_{2}$ are restricted to the time interval $(t',t)$, and the corresponding $\rho_{\bf k}^{(2)><}(\nu)$ is also readily obtainable from the $G_{0}W_{0}$ self-energy contained in  Fig. \ref{Sigma_k}c. Rather, the problem arises in that the analogous $W$-induced fluctuations across the Fermi surface exist also outside this interval and hence in a {\it consistent} approach should be coupled to the quasiparticle injected into the same region of the phase space. This leads to the standard, adiabatic formulation of Green's function (\ref{eq:causalGN}) amenable to usual perturbation treatment which, in general, is not representable in cumulant form [\onlinecite{Doniach}].  Analogous argument holds also for holes created near the Fermi surface.
 However, cumulant representation with {\it partly restored afore required consistency} is regained by resorting to a hybrid approach in which the  renormalization of one-particle states in the intervals $(-\infty,t')$ and $(t,\infty)$ is accounted for through their representation by stationary KS states, and within the excitation interval $(t',t)$ through {\it dynamical} renormalization by the generically same interaction $W$ using cumulant expansion. The second order cumulant is in this case generated by the full $G_{0}W_{0}$ self-energy, as implemented in [\onlinecite{AHK}] and extensively used thereafter. However, the degree of consistency of this matching  near the Fermi level is not {\it a priory} clear and deviations from the standard Feynman-Dyson perturbation expansion remain to be explored.

\section{Temporal evolution of the amplitude and phase of the quasiparticle propagator}
\label{sec:L}

Using the closed form solution (\ref{eq:Gcumul}) for the propagator in cumulant representation one can introduce the quantities which conveniently measure temporal evolution of the quasiparticle upon its promotion into the initial  state $|{\bf k}\rangle$. Again we first discuss the case of electron propagators and spectra and then deduce the corresponding quantities for holes. 
In this context the quantity of primary interest is the survival probability of the quasiparticle initial state
\bq
L_{\bf k}(t)=\mid G_{\bf k}(t)\mid^2=\exp[2\mbox{Re}C_{\bf k}(t)],
\label{eq:L}
\eq
and the evolution of its phase defined by
\bq
\phi_{\bf k}(t)=-\mbox{Im}\ln\left[iG_{\bf k}(t)\right]=\epsilon^{0}_{\bf k}t-\mbox{Im}C_{\bf k}(t)=\epsilon^{0}_{\bf k}t+\varphi_{\bf k}(t).
\label{eq:phi}
\eq
In this notation the derivative $\partial \varphi_{\bf k}(t)/\partial t$ describes the relaxation of quasiparticle energy in the course of time. 

Of particular interest is the behavior of (\ref{eq:L}) and (\ref{eq:phi}) on the various time scales characteristic of the studied system. 
Quite generally, the quasiparticle evolution generated by the Hamiltonian (\ref{eq:H}) and described by (\ref{eq:L}) and (\ref{eq:phi})  exhibits three distinct consecutive stages [\onlinecite{PSS2012}]:\\
(i) early time non-Markovian evolution characterized by the initial off-the-energy-shell transients,\\
(ii) Intermediate stage quasi-stationary Markovian evolution characterized by the exponential decay of quasiparticle probability amplitude (\ref{eq:L}) and stationary derivative of the phase $\partial \varphi_{\bf k}(t)/\partial t$, \\
 (iii) Collapse of the quasiparticle amplitude and phase in the asymptotic limit $t\rightarrow\infty$. \\
The remainder of this section discusses the characteristics of these evolution regimes.

\subsection{Initial transients and the crossover to quasistationary regime of quasiparticle propagation}
\label{sec:Transients}

The early quasiparticle propagation past its injection into the eigenstate $|{\bf k}\rangle$ of $H^0$ is before the establishment of energy conservation governed by initial off-the-energy-shell transients. The first transient is the Zeno decay 
\bq
L_{\bf k}(t\rightarrow 0)=\exp(-t^2/\tau_Z^2),
\label{eq:LZeno}
\eq
where $\tau_{Z}$ is the so called Zeno time which measures the convexity of the initial drop of the amplitude of the quasiparticle during its earliest ballistic propagation in the band [\onlinecite{Pascazio}]. This is followed by virtual excitation of the modes that constitute the response of the heatbath described by $H^0_{bos}$. Since the coupling to coherent collective modes of the heatbath like plasmons is usually strong, the initial transients in (\ref{eq:L}) and (\ref{eq:phi}) may be dominated by oscillations with the period of inverse plasmon frequency. The onset of steady state quasiparticle propagation is governed by the establishment of energy conservation; this happens when the functionals $(1-\cos\nu t)/\nu^2$ and $\sin\nu t/\nu$ in the general integral representation of $C_{\bf k}(t)$ defined by (\ref{eq:C_rho}) can be replaced by their equivalent long time asymptotic forms $\pi\delta(\nu)|t|$ and $\pi\delta(\nu)$, respectively. Since this strongly depends on the structure of $\rho_{\bf k}(\nu)$ the onsets are very system specific. Illustrative examples of such specifities pertaining to holes excited in Q2D bands on Ag(111) and Cu(111) surfaces are presented in Figs. 4 and 5 of Ref. [\onlinecite{SGLD}]. To grasp their emergence it is of utmost importance to note that the description of establishment of steady state propagation does not require the introduction of an additional infinitesimal $i\eta$ into the denominator of the integrand on the RHS of (\ref{eq:C_rho}) as long as the oscillating terms are retained. Namely, the neglect of oscillatory terms and simultaneous introduction of $i\eta$ in the denominator is an auxiliary procedure in the treatments based on the assumption of {\it adiabatic} temporal boundary conditions. The latter are typical of the scattering experiments in which the projectile-target interaction is switched on and off adiabatically whereas in the present  problem we deal with the instantaneous switching on of $\lambda V$ that is typical of photoemission boundary conditions. It is only in the long time or asymptotic limit that the two types of temporal boundary conditions may give the same results.  {\it In other words, the kernel of the integral transform on the RHS of (\ref{eq:C_rho}), viz. the function $(1-e^{-i\nu t}-i\nu t)/\nu^2$, takes a proper account of the passage of the quasiparticle from the initial transient, energy non-conserving interval,  to subsequent quasi-stationary regime of propagation in which the energy of the quasiparticle-boson system is conserved.}

\subsection{Quasistationary regime}
\label{sec:Qstationary}

\subsubsection{Polarization energy shift}
\label{sec:relshift}

Once the quasi-stationary regime is reached one can single out from $C_{\bf k}(t)$ its imaginary part linear in $t$ that determines the long time behavior of $\varphi_{\bf k}(t)$ in (\ref{eq:phi}). This  gives the polarization or energy relaxation shift $\Delta_{\bf k}$ of the unperturbed level energy $\epsilon^{0}_{\bf k}$, viz.
\bq
C^{\rm{pol}\stackrel{>}{\scriptscriptstyle <}}_{\bf k}(t)=- i\left[\mp\int d\nu \frac{\rho_{\bf k}^{\stackrel{>}{\scriptscriptstyle <}}(\nu)}{\nu}\right]t=-i\Delta_{\bf k}^{\stackrel{>}{\scriptscriptstyle <}}t.
\label{eq:Crel}
\eq
The integral is taken in the sense of principal value with the range of integration extended over the whole bandwidth of $\rho_{\bf k}^{\stackrel{>}{\scriptscriptstyle <}}(\nu)$. This is in accord with the notion of polarization or relaxation energy involving all virtual, off-the-energy shell excitations embodied in $\rho_{\bf k}^{\stackrel{>}{\scriptscriptstyle <}}(\nu)$. Thus, substituting the leading cumulant joint density of excitations $\rho_{\bf k}^{(2)\stackrel{>}{\scriptscriptstyle <}}(\nu)$ given by (\ref{eq:rho2>}) and (\ref{eq:rho2<}) into expression (\ref{eq:Crel}) one obtains the second order Rayleigh-Schr\"{o}dinger (RS) perturbation theory correction  to the unperturbed quasiparticle energy $\epsilon^{0}_{\bf k}$ derived from expression (\ref{eq:Sigma2k}), viz.
\bq
\Delta_{\bf k}^{(2)\stackrel{>}{\scriptscriptstyle <}}=\mbox{Re} \Sigma_{\bf k}^{(2)\stackrel{>}{\scriptscriptstyle <}}(\epsilon_{\bf k}^{0})\propto \lambda^{2},
\label{eq:Delta}
\eq
 and analogously so for the fourth order terms $\propto\lambda^{4}$ [\onlinecite{Mahan_book}].  This illustrates that the natural basis for cumulant representation (\ref{eq:Gcumul}) is the space spanned by the one-particle eigenstates and eigenenergies of $H^{0}$ [\onlinecite{Kubo,Mahan_book}], and  not by any partially renormalized quantities (e.g. like those derived from SC Brillouin-Wigner perturbation theory), as that would lead to {\it overcounting} of the terms of the same powers in $\lambda$. This could have been inferred also from the explicit form of  the second order cumulant (\ref{eq:C2}).

\subsubsection{Quasistationary propagation of recoiling quasiparticles}
\label{sec:recoil}

The cumulant function $C_{\bf k}(t)$ may embody another contribution linear in $t$ besides the purely imaginary term  (\ref{eq:Crel}) yielding the polarization energy. This arises from the long time limit of $(1-\cos\nu t)/\nu^2\rightarrow \pi\delta(\nu)|t|$ in the integral representation of $C_{\bf k}(t)$ [cf. expression (\ref{eq:C_rho})], which is nonvanishing provided there exists a quasiparticle component $\rho_{\bf k}^{\rm qp}(\nu)$ of $\rho_{\bf k}^{\stackrel{>}{\scriptscriptstyle <}}(\nu)$ with smooth low energy behaviour
\bq
\rho_{\bf k}^{\rm qp}(\nu\rightarrow 0)=\rho_{\bf k}^{\rm qp}(0)+\nu \partial \rho_{\bf k}^{\rm qp}(\nu)/\partial\nu |_{\nu=0}.
\label{eq:rhoqp0}
\eq
Such a quasistationary property of $\rho_{\bf k}^{\rm qp}(\nu)$ requires that the quasiparticle recoil energy $\epsilon_{\bf k-q}^{0}-\epsilon_{\bf k}^{0}$ be degenerate with the continuum of heatbath excitations like the electron-hole pairs, acoustic phonons, etc. This yields the long time behavior of the quasiparticle component $C_{\bf k}^{\rm qp}(t)$ of $C_{\bf k}(t)$ comprising the terms linear in $t$ and a complex constant, viz.
\bq
\lim_{t\rightarrow\pm\infty} C_{\bf k}^{{\rm qp}\stackrel{>}{\scriptscriptstyle <}}(t)= -i(\Delta_{\bf k}^{{\rm qp}\stackrel{>}{\scriptscriptstyle <}} \mp i\Gamma_{\bf k}^{{\rm qp}\stackrel{>}{\scriptscriptstyle <}})t-w_{\bf k}^{{\rm qp}\stackrel{>}{\scriptscriptstyle <}}.
\label{eq:Clarge}
\eq
Here
\bq
\Gamma_{\bf k}^{{\rm qp}\stackrel{>}{\scriptscriptstyle <}}=\pi\rho_{\bf k}^{{\rm qp}\stackrel{>}{\scriptscriptstyle <}}(0),
\label{eq:Gamma}
\eq
has the meaning of the quasiparticle decay rate per unit time and is an on-the-energy-shell quantity, in contrast to the polarization energy $\Delta_{\bf k}^{{\rm qp}\stackrel{>}{\scriptscriptstyle <}}$.  The structure of the complex constant $w_{\bf k}^{{\rm qp}\stackrel{>}{\scriptscriptstyle <}}$ which derives from quasistationary $\rho_{\bf k}^{{\rm qp}\stackrel{>}{\scriptscriptstyle <}}(\nu)$ around the excitation threshold  $\nu=0$ can be deduced by using the equivalence of long time limits of the kernel of integral transformation (\ref{eq:C_rho}) and its stationary representation
\barr
w_{\bf k}^{{\rm qp}\stackrel{>}{\scriptscriptstyle <}}&=&\lim_{t\rightarrow\pm\infty} \int d\nu \frac{1-e^{\mp i\nu t}}{\nu^2}\rho_{\bf k}^{{\rm qp}\stackrel{>}{\scriptscriptstyle <}}(\nu)\rightarrow  \int d\nu \frac{\rho_{\bf k}^{{\rm qp}\stackrel{>}{\scriptscriptstyle <}}(\nu)}{(\nu\pm i\eta)^2}\nonumber\\
&=&
 -\int d\nu \rho_{\bf k}^{{\rm qp}\stackrel{>}{\scriptscriptstyle <}}(\nu)\frac{\partial}{\partial\nu}\frac{1}{\nu
\pm i\eta}= \int d\nu \frac{\frac{\partial}{\partial\nu}\rho_{\bf k}^{{\rm qp}\stackrel{>}{\scriptscriptstyle <}}(\nu)}{\nu \pm i\eta}
\label{eq:limC}
\earr
In the case of second order cumulants we substitute expressions (\ref{eq:rho2>}) and (\ref{eq:rho2<}) into  (\ref{eq:Crel}) and (\ref{eq:limC}),  change the integration variables and use the Kramers-Kronig relations to obtain
\barr
\Delta_{\bf k}^{{\rm qp}\stackrel{>}{\scriptscriptstyle <}}\mp i\Gamma_{\bf k}^{{\rm qp}\stackrel{>}{\scriptscriptstyle <}}&=&\mbox{Re}\Sigma_{\bf k}^{{\rm qp}\stackrel{>}{\scriptscriptstyle <}}(\epsilon_{\bf k}^{0})+i\mbox{Im}\Sigma_{\bf k}^{{\rm qp}\stackrel{>}{\scriptscriptstyle <}}(\epsilon_{\bf k}^{0}),
\label{eq:onshellSigma2}\\
 w_{\bf k}^{{\rm qp}\stackrel{>}{\scriptscriptstyle <}}&=&-\frac{\partial}{\partial\nu}\Sigma_{\bf k}^{{\rm qp}\stackrel{>}{\scriptscriptstyle <}}(\nu)|_{\nu=\epsilon_{\bf k}^{0}}.
\label{eq:w2}
\earr
Relations (\ref{eq:Clarge}), (\ref{eq:onshellSigma2}) and (\ref{eq:w2}) signify that the analytic properties of $G_{\bf k}^{\rm qp}(\omega)$ deriving from (\ref{eq:Clarge}) are determined dominantly by the pole located at $\omega=\epsilon^{0}_{\bf k}+\Delta_{\bf k}^{{\rm qp}\stackrel{>}{\scriptscriptstyle <}}\mp i\Gamma_{\bf k}^{{\rm qp}\stackrel{>}{\scriptscriptstyle <}}$ in the complex plane, with the weight (residue) $Z_{\bf k}=\exp(-w_{\bf k}^{{\rm qp}\stackrel{>}{\scriptscriptstyle <}})$. Note that other nonanalytic structures that may appear in the full $G_{\bf k}(\omega)$, e.g. like those associated with the cuts in complex plane, give rise to different temporal asymptotic behavior of $G_{\bf k}(t)$ that {\it can not be represented by the simple form} (\ref{eq:Clarge}) (see subsection \ref{sec:recoilless}).

For later convenience we may generalize these results to several bands. This implies that all intermediate state summations should also include allowed interband transitions $({\bf k},n)\rightarrow ({\bf k'},n')$. Then, the asymptotic behavior of the cumulants (\ref{eq:Clarge}) and (\ref{eq:C1}) produces a peak in the quasiparticle spectrum in the $n$th band described by 
\bq
{\cal N}_{{\bf k},n}^{{\rm qp}\stackrel{>}{\scriptscriptstyle <}}(\omega)=
\frac{\left(\stackrel{1-n_{{\bf k},n}}{\scriptscriptstyle n_{{\bf k},n}}\right)}{\pi}
e^{-w_{{\bf k},n}}
\frac{\left[\Gamma_{{\bf k},n}\cos \alpha_{{\bf k},n}-(\omega-E_{{\bf k},n})\sin\alpha_{{\bf k},n}\right]}{(\omega-E_{{\bf k},n})^{2}+\Gamma_{{\bf k},n}^{2}}
\label{eq:Nqp}
\eq
 where in the shorthand notation $w_{{\bf k},n}=\mbox{Re}w_{{\bf k},n}^{{\rm qp}\stackrel{>}{\scriptscriptstyle <}}$, $\alpha_{{\bf k},n}=\mbox{Im}w_{{\bf k},n}^{{\rm qp}\stackrel{>}{\scriptscriptstyle <}}$, $E_{{\bf k},n}=\epsilon_{{\bf k},n}^{0}+\Delta_{{\bf k},n}^{{\rm qp}\stackrel{>}{\scriptscriptstyle <}}-v_{{\bf k},n}^{xc}$, and $\Gamma_{{\bf k},n}=\Gamma_{{\bf k},n}^{{\rm qp}\stackrel{>}{\scriptscriptstyle <}}$.  The weight of (\ref{eq:Nqp}) is reduced by the Debye-Waller factor-like expression $\exp(-w_{{\bf k},n})$, whereas its nonvanishing imaginary part $\alpha_{{\bf k},n}$ gives rise to deviations of the spectral shape from a pure Lorentzian. The remaining spectral weight  is shifted to the inelastic wings and satellite structure. Note also that contrary to the popular belief, owing to the short time behaviour (\ref{eq:LZeno}) the wings of the spectrum far from the quasiparticle energy $E_{{\bf k},n}$ do not acquire a Lorentzian lineshape. Moreover, in the case of electron promotion into the lowest unoccupied (or of hole into the highest occupied) band state  $|{{\bf k},n}\rangle$ the threshold peak of the quasiparticle spectrum reduces to a $\delta$-function followed by the one-sided inelastic wing [\onlinecite{MHRT,PSS2012}]. 

Exponential decay of the quasiparticle amplitude (\ref{eq:Clarge}) cannot continue indefinitely in bands of finite width.  Once the electron (hole) reaches the lowest unoccupied (highest occupied) level its evolution collapses from exponential to power law asymptotic decay [\onlinecite{Khalfin}]. This is illustrated in Fig. 2 of Ref. [\onlinecite{Pastawski}] and Fig. 10 of Ref. [\onlinecite{PSS2012}]. 

\subsection{Temporal evolution of recoiless quasiparticles}
\label{sec:recoilless}

For the behavior of $\rho_{\bf k}(\nu)$ around $\nu=0$ that is different from that outlined in subsection \ref{sec:recoil}, the long time limit of $C_{\bf k}(t)$ may acquire a nonlinear time dependence (cf. Sec. 3 of Ref. [\onlinecite{PSS2012}]). This, in turn, produces a profound effect on the quasiparticle spectrum (\ref{eq:N}). The most familiar case is the forward scattering limit of recoilless quasiparticles arising from their infinite effective mass in flat bands so that $\epsilon_{\bf k-q}^{0}-\epsilon_{\bf k}^{0}\simeq 0$. A  classical example of such a reduction of the Hamiltonian (\ref{eq:H}) is provided by a core hole coupled to the continuum of incoherent e-h excitations constituting the system  heatbath [\onlinecite{Langreth,PSS1984}]. In this case the low energy or infra red (IR) limit of the joint density of states (\ref{eq:rho}) takes the form [\onlinecite{MHRT}]
\bq
\rho_{\rm IR}(\nu) = \nu\left.\frac{\partial \rho}{\partial\nu}\right|_{\nu=0}=\alpha\nu,   \hspace{5mm} \nu\leq \Omega_{\rm IR},
\label{eq:rhoIR}
\eq
where due to the infinite mass of the source the subscript ${\bf k}$ is omitted and 
$\Omega_{\rm IR}$ denotes the cut-off parameter for the linear dependence in  (\ref{eq:rhoIR}). Taking for calculational convenience the exponential form of the cut-off, $\exp(-\nu/\Omega_{\rm IR})$, we obtain an exact result 
\bq
C_{\rm IR}(t)=-i(\mp\alpha\Omega_{\rm IR})t-\alpha\ln(1\pm i\Omega_{\rm IR} t).
\label{eq:CIR}
\eq
where again the upper and lower signs refer to electrons (positive $t$) and holes (negative $t$), respectively. Substitution of (\ref{eq:CIR}) into (\ref{eq:Gcumul}) leads to the quasiparticle propagator
\bq
G_{\rm IR}(t)=\mp i \theta(\pm t) e^{-i\epsilon^{0}t}\frac{e^{\pm i\alpha\Omega_{\rm IR}t}}{(1\pm i\Omega_{\rm IR}t)^{\alpha}}.
\label{eq:GIR}
\eq
This exhibits the correct short time limit (\ref{eq:LZeno}) which is directly succeeded by the long time limit of asymptotic power law decay $\propto 1/t^{\alpha}$, i.e. without the passage through the stage of exponential decay $\propto \exp(-\Gamma t)$ characteristic of the recoiling quasiparticle regime described by Eq. (\ref{eq:Clarge}). This is due to the absence of pole(s) from the FT of (\ref{eq:GIR}), whereby its (non)analitical properties arise exclusively from the cut along the positive (negative) real axis in the $\omega$-plane. This renders the spectrum exhibiting the IR power law divergence at the excitation threshold
\bq
{\cal N}_{\rm IR}^{\stackrel{>}{\scriptscriptstyle <}}(\omega)=\frac{e^{-(\omega-\epsilon^{0}\pm \alpha\Omega_{\rm IR})/\Omega_{\rm IR}}\theta(\omega-\epsilon^{0}\pm \alpha\Omega_{\rm IR})}{\Omega_{\rm IR}^{\alpha}\Gamma(\alpha)(\omega-\epsilon^{0}\pm \alpha\Omega_{\rm IR})^{1-\alpha}},
\label{eq:NIR}
\eq
where $\Gamma(\alpha)$ is the Gamma function, and $\alpha$ plays the role of critical exponent of the IR power law divergence.  A peculiar property of the spectrum (\ref{eq:NIR}) in comparison with (\ref{eq:Nqp}) is the complete suppression of the elastic component at the quasiparticle threshold excitation energy (since the corresponding Debye-Waller exponent is divergent, i.e. $\mbox{Re}w_{\rm IR}\rightarrow \infty$ for $t\rightarrow\infty$), and the removal of the entire spectral weight to the IR divergent inelastic sidewing.

\subsection{Satellite structure in the quasiparticle spectrum}
\label{sec:satellites}

The joint density of excitations $\rho_{\bf k}(\nu)$ may embody, besides the  quasiparticle component $\rho_{\bf k}^{\rm qp}(\nu)$ that is continuous around $\nu=0$ and leads to (\ref{eq:Clarge}), also the components that exhibit sharp peaks located at $\omega^{\rm sat}$ away from $\nu=0$ (cf. Fig. 3 in Ref. [\onlinecite{McMullen}]), viz. that we have
\bq
\rho_{\bf k}(\nu)=\rho_{\bf k}^{\rm qp}(\nu)+\sum_{\rm sat}\rho_{\bf k}^{\rm sat}(\nu).
\label{eq:rho_qp+sat}
\eq 
In electron gas such components $\rho_{\bf k}^{\rm sat}(\nu)$ may originate from plasmon modes of frequency $\omega_{\bf q}^{\rm pl}$. A prerequisite for the occurrence of a prominent peak in $\rho_{\bf k}^{\rm sat}(\nu)$ around $\nu=\omega^{\rm sat}$ is the small quasiparticle recoil energy $\epsilon_{\bf k}^{0}- \epsilon_{\bf k-q}^{0}\ll\omega_{\bf q}^{\rm pl}\simeq \omega^{sat}$.
This gives rise to strong contributions from the oscillating term in the integrand on the RHS of (\ref{eq:C_rho2}), the so called satellite generator, and produces contributions to $C_{\bf k}(t)$ oscillating with $\omega^{\rm sat}$. 

\subsubsection{Satellites in the plasmonic polaron model}
\label{sec:polaron}

To demonstrate the occurrence of satellite structures we first introduce the simple plasmonic polaron model based on the ansatz for plasmon pole(s) dominated ${\cal D}_{\bf q}^{\rm pl}(\omega')=\frac{1}{\pi}|\mbox{Im}\chi_{\bf q}^{\rm pl}(\omega')|={\cal S}_{\bf q}^{\rm pl}\delta(\omega'-\omega_{\bf q}^{\rm pl})$. Here ${\cal S}_{\bf q}^{\rm pl}$ and $\omega_{\bf q}^{\rm pl}$ are the weight (residue) of a particular plasmon pole. In the limit of small recoil this leads to 
\barr
\rho_{\bf k}^{{\rm sat}\stackrel{>}{\scriptscriptstyle <}}(\nu)&=& \sum_{\bf q}v_{\bf k-q,k}^2 {\cal S}_{\bf q}^{\rm pl}\delta(\nu-(\omega_{\bf q}^{\rm pl}\pm \epsilon_{\bf k-q}^{0}\mp \epsilon_{\bf k}^{0}))\nonumber\\
&\simeq&
 w_{\bf k}^{\rm sat}(\omega_{\bf k}^{\rm sat})^{2}\delta(\nu-\omega_{\bf k}^{\rm sat})
\label{eq:rho_sat}
\earr
 in which the thus defined $w_{\bf k}^{\rm sat}$  is generally strongly ${\bf k}$-dependent, whereas the ${\bf k}$-dependence of $\omega_{\bf k}^{\rm sat}$ is expected to be weak, i.e. $\omega_{\bf k}^{\rm sat}\simeq \omega^{\rm sat}$ as it originates from  weakly dispersive $\omega_{\bf q}^{\rm pl}$. Substituting this into (\ref{eq:C_rho}) we obtain
\bq
C_{\bf k}^{{\rm sat}\stackrel{>}{\scriptscriptstyle <}}(t)=-i\Delta_{\bf k}^{{\rm sat}\stackrel{>}{\scriptscriptstyle <}}t + w_{\bf k}^{{\rm sat}\stackrel{>}{\scriptscriptstyle <}}e^{\mp i\omega^{\rm sat}t}-w_{\bf k}^{{\rm sat}\stackrel{>}{\scriptscriptstyle <}}
\label{eq:Csat}
\eq
where 
\barr
\Delta_{\bf k}^{{\rm sat}\stackrel{>}{\scriptscriptstyle <}}&=& \Delta_{\bf k}^{{\rm sat}\stackrel{>}{\scriptscriptstyle <}}(\epsilon_{\bf k}^{0})\simeq \mp \sum_{\bf q}{\cal S}_{\bf q}^{\rm pl}v_{\bf k-q,k}^{2}/\omega^{sat},  
\label{eq:Delta_sat}\\
  \Gamma_{\bf k}^{{\rm sat}\stackrel{>}{\scriptscriptstyle <}}&=& \Gamma_{\bf k}^{{\rm sat}\stackrel{>}{\scriptscriptstyle <}}(\epsilon_{\bf k}^{0})=0,
\label{eq:Gamma_sat}\\
w_{\bf k}^{{\rm sat}\stackrel{>}{\scriptscriptstyle <}}&\simeq& \sum_{\bf q}{\cal S}_{\bf q}^{\rm pl}\left(\frac{v_{\bf k-q,k}}{\omega^{sat}}\right)^{2}=
\frac{\left|\Delta_{\bf k}^{{\rm sat}\stackrel{>}{\scriptscriptstyle <}}\right|}{\omega^{\rm sat}},
\label{eq:w_sat}
\earr
 Thereby each source peak in $\rho_{\bf k}^{\rm sat}(\nu)$ gives rise to the corresponding energy shift (\ref{eq:Delta_sat}), the satellite generator $w_{\bf k}^{\rm sat}e^{\mp i\omega^{\rm sat}t}$, and the DWE (\ref{eq:w_sat}), and they are all additive in $C_{\bf k}(t)$ and hence multiplicative in (\ref{eq:Gcumul}).
Consequently, the spectrum (\ref{eq:N}) obtained from the FT of the product of $e^{-w_{\bf k}^{\rm sat}}$ and the expanded exponential $\exp(w_{\bf k}^{\rm sat}e^{\mp i\omega^{\rm sat}t})$  acquires the form of a series of discernible  maxima or satellites located at integer multiples $\mp l\omega^{\rm sat}$ of the fundamental satellite frequency and weighted by $e^{-w_{\bf k}^{\rm sat}}(w_{\bf k}^{\rm sat})^{l}/l!$. Their shapes are given by the convolution of the narrow satellite generating peak with the quasiparticle peak given by (\ref{eq:Nqp}) or (\ref{eq:NIR}). In other words,  each satellite peak in the spectrum (\ref{eq:N}) replicates the convolution with the structure of the quasielastic threshold peak [\onlinecite{Langreth,AHK}].
Therefore, the quasiparticle spectrum in the plasmonic polaron model can be represented by a generic multiboson excitation expression frequently quoted in the literature [\onlinecite{Langreth,Dunn,PSS1984,Mahan_book,recovery,Reining,Mishchenko}] (for generality we restore the band index $n$)

\begin{figure}[tb]
\rotatebox{0}{\epsfxsize=8.5cm \epsffile{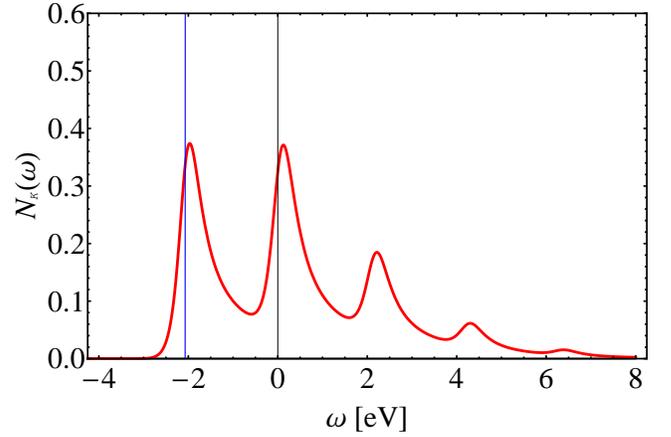} }
\caption{(Color online) Illustration of the quasiparticle spectrum ${\cal N}_{\bf K}(\omega)$  corresponding to propagation of a hot electron in an initially empty Q2D surface band completely above the Fermi level. The electron is coupled to the electronic charge density excitations in the underlying substrate whose spectrum consists of incoherent e-h pairs and surface plasmons of energy  $\hbar\omega_{s}=2$ eV.  The calculation is performed for intermediate coupling strength  and initial 2D electron momentum $K=0.06$ a.u.,  which produces nearly equal strengths of the threshold and first satellite peak. The satellites appear approximately at multiples of the surface plasmon energy above the threshold energy $\epsilon_{\bf K}=\epsilon_{\bf K}^{0}+\Delta_{\bf K}$  of the leftmost peak (denoted by the vertical  line at $\omega\simeq -2$ eV). Energy zero corresponds to the first moment of the spectrum which measures the energy shift $\Delta_{\bf K}<0$ of the unperturbed level $\epsilon_{\bf K}^{0}$. }
\label{RescaledSpectrum}
\end{figure}

\bq
{\cal N}_{{\bf k}}^{\stackrel{>}{\scriptscriptstyle <}}(\omega)=\sum_{n, l=0}^{\infty}e^{-w_{{\bf k},n}^{\rm sat}}\frac{(w_{{\bf k},n}^{\rm sat})^{l}}{l!}{\cal N}_{{\bf k},n}^{{\rm qp}\stackrel{>}{\scriptscriptstyle <}}(\omega-\epsilon_{{\bf k},n}^{0}-\Delta_{{\bf k},n}\mp l\omega^{\rm sat}),
\label{eq:Npolaron}
\eq
where ${\cal N}_{{\bf k},n}^{{\rm qp}\stackrel{>}{\scriptscriptstyle <}}(\omega)$ is given by (\ref{eq:Nqp}) and 
\bq
\Delta_{{\bf k},n}=\Delta_{{\bf k},n}^{\rm qp}+\Delta_{{\bf k},n}^{\rm sat},
\label{eq:Delta_tot}
\eq
is determined from (\ref{eq:onshellSigma2}) and (\ref{eq:Delta_sat}) or from a single calculation (\ref{eq:Crel}). 
 The spectrum  described by (\ref{eq:Npolaron}) follows the general pattern shown in Fig. \ref{RescaledSpectrum}. A completely analogous formula applies also in the limit of recoilles quasiparticles in which case ${\cal N}^{\rm qp}(\omega)$ is replaced by ${\cal N}_{IR}(\omega)$ [\onlinecite{Langreth,PSS1984,PSS2012}]. Note that in the opposite limit in which the quasiparticle recoil blurs the  plasmon induced satellite in $\rho_{{\bf k},n}^{\rm sat}(\nu)$ and gives rise to $\Gamma_{\bf k}^{{\rm sat}\stackrel{>}{\scriptscriptstyle <}}(\epsilon_{\bf k}^{0})>0$ the regime described by Eq. (\ref{eq:Clarge}) is also applicable to descriptions of excitation of real plasmons by the propagating quasiparticle (cf. Ref. [\onlinecite{GN1975}] and Figs. 3 and 4 in Ref. [\onlinecite{4cumul}]).

\subsubsection{Satellites in ab initio GW+C approach} 
\label{sec:SatelliteGW+C}

The situation becomes more complex when instead of the plasmon pole ansatz (\ref{eq:rho_sat}) the corresponding GW self-energy derived from first principles is used. In this case the satellite generating component of $C_{{\bf k},n}^{(2)\stackrel{>}{\scriptscriptstyle <}}(t)$ which avoids overcounting of contributions leading to (\ref{eq:Clarge}) reads
\barr
C_{{\bf k},n}^{{\rm sat}\stackrel{>}{\scriptscriptstyle <}}(t)&=&- \int d\nu \frac{[\rho_{{\bf k},n}^{(2)\stackrel{>}{\scriptscriptstyle <}}(\nu)-\rho_{{\bf k},n}^{(2){\rm qp}\stackrel{>}{\scriptscriptstyle <}}(\nu)]}{\nu^2} (1-e^{\mp i\nu t})\nonumber\\
&=&
-w_{{\bf k},n}^{{\rm sat}\stackrel{>}{\scriptscriptstyle <}}+{\cal S}_{{\bf k},n}^{{\rm sat}\stackrel{>}{\scriptscriptstyle <}}(t),
\label{eq:CsatGW}
\earr
where $\rho_{{\bf k},n}^{(2){\rm qp}\stackrel{>}{\scriptscriptstyle <}}(\nu)$ is approximated with (\ref{eq:rhoqp0}). Here the $l$th satellite in the quasiparticle spectrum is generated by the $l$th term in the expansion of $\exp[{\cal S}_{{\bf k},n}^{{\rm sat}\stackrel{>}{\scriptscriptstyle <}}(t)]$ in a power series. This requires calculations of the $l$-fold convolutions of the FT of ${\cal S}_{{\bf k},n}^{{\rm sat}\stackrel{>}{\scriptscriptstyle <}}(t)$ with the quasiparticle peak (\ref{eq:Nqp}), and this is practically intractable for higher $l$s. However, since in the majority of systems of interest the quasiparticle coupling to plasmons is relatively weak (as measured by the effective $w_{{\bf k},n}^{\rm sat}<1$ in practical applications only the first satellite need be computed. This is rather advantageous because the FT of ${\cal S}_{{\bf k},n}^{{\rm sat}\stackrel{>}{\scriptscriptstyle <}}(t)$ is obtained in a simple form [\onlinecite{AHK}]
\bq
{\cal N}_{{\bf k},n}^{{\rm sat}\stackrel{>}{\scriptscriptstyle <}}(\omega)=\frac{\beta_{{\bf k},n}^{\stackrel{>}{\scriptscriptstyle <}}(\omega)-\left(\beta_{{\bf k},n}^{\stackrel{>}{\scriptscriptstyle <}}(\epsilon_{{\bf k},n}^{0})+(\omega-\epsilon_{{\bf k},n}^{0})\frac{\partial}{\partial\omega}\beta_{{\bf k},n}^{\stackrel{>}{\scriptscriptstyle <}}(\epsilon_{{\bf k},n}^{0})\right)}{(\omega-\epsilon_{{\bf k},n}^{0})^2}
\label{calComega}
\eq
where $\beta_{{\bf k},n}^{\stackrel{>}{\scriptscriptstyle <}}(\omega)=\mp\frac{1}{\pi}\mbox{Im}\Sigma_{{\bf k},n}^{(2)\stackrel{>}{\scriptscriptstyle <}}(\omega)$. Hence, the quasiparticle spectrum obtained by combining the {\it ab initio} $G_{0}W_{0}$ selfenergy and cumulant approach takes the form %
\barr
{\cal N}_{{\bf k}}^{\stackrel{>}{\scriptscriptstyle <}}(\omega)&=&\sum_{n}e^{-w_{{\bf k},n}^{{\rm sat}\stackrel{>}{\scriptscriptstyle <}}}[{\cal N}_{{\bf k},n}^{{\rm qp}\stackrel{>}{\scriptscriptstyle <}}(\omega)+{\cal N}_{{\bf k},n}^{{\rm qp}\stackrel{>}{\scriptscriptstyle <}}(\omega)\ast{\cal N}_{{\bf k},n}^{{\rm sat}\stackrel{>}{\scriptscriptstyle <}}(\omega)\nonumber\\
&+&
 {\cal O}((w_{{\bf k},n}^{{\rm sat}\stackrel{>}{\scriptscriptstyle <}})^{2})],
\label{eq:Nsat1}
\earr
where $\ast$ denotes convolution.
The quantitative difference between the spectra (\ref{eq:Npolaron}) and (\ref{eq:Nsat1}) computed for a concrete system is illustrated in Sec. \ref{sec:Si}.

\section{Connection of cumulant expansion with GW approximations}
\label{sec:GWlink}

Using expressions derived in Secs. \ref{sec:propagators_cumul} and \ref{sec:L} we are now in a position to assess the desired connection between the standard self-energy and cumulant expansion of the single particle propagator (\ref{eq:causalG}) whose time evolution is governed by the dynamic interaction $W$ (\ref{eq:W}). To this end we consider the regime of small but finite particle recoil in which the analytic properties of the renormalized propagator (\ref{eq:GSigma}) are dominated by isolated simple poles, with only small contribution(s) from the cut(s). Thereby we avoid the limit (\ref{eq:CIR}) and allow for the appearance of distinct plasmon-generated peaks in the quasiparticle spectra. In this case we find that the long time behavior of the second order cumulant, which dominantly determines the quasiparticle spectral properties,  is given by expressions (\ref{eq:Clarge}) and (\ref{eq:CsatGW}). Their gross features should according to (\ref{eq:C_rho}) persist also in higher order cumulants.  This means that the temporal properties of the cumulant series, and thereby of ensuing $G_{\bf k}(t)$, are already determined by the $G_{0}W_{0}$-generated second order cumulant (\ref{eq:C_rho2}).   As pointed out in Ref. [\onlinecite{Mahan_book}] this usually yields the best results for the physics of unbound polaron problem described by (\ref{eq:H}).

Now, in view of the same form of integral representation of the general cumulant expression (\ref{eq:C_rho}) and of the second order term (\ref{eq:C_rho2}) it would be tempting to extrapolate the second order result to the whole cumulant series (\ref{eq:Cl}) and replace it by expressions calculated from some higher order terms of the GW self-energy $\Sigma_{{\bf k},n}^{GW}(\omega)$.  However, for the present problem defined by (\ref{eq:H}) this procedure is not justified and should be avoided. This is easily verified by examining the explicit forms of the next higher, fourth order cumulants $C_{{\bf k},n}^{(4)}(t)\propto \lambda^{4}$ that have been calculated for the present model [\onlinecite{Dunn,4cumul}].  The corresponding diagrams encompass the contributions with two noncrossing boson lines  (GWA generic), two crossing boson lines (vertex corrections), and a product of two uncorrelated second order corrections in the time domain (cf. Fig. 6.4 in Ref. [\onlinecite{Mahan_book}]).  All these terms are needed to construct the correct $C_{{\bf k},n}^{(4)}(t)$ whose long time limit encompasses the on-the-energy-shell fourth order Rayleigh-Schr\"{o}dinger self-energy terms rather than solely the GWA-generated self-energy terms [\onlinecite{Mahan_book}]. The same conclusion regarding the noncrossing and crossing boson line diagrams applies to higher order cumulants [\onlinecite{Kubo}], and thereby also to the whole cumulant series (\ref{eq:Cl}).
This means that the replacement of the long time limit of $C_{{\bf k},n}(t)$ in (\ref{eq:Cl}) by expressions deriving from selfconsistent or higher order $\Sigma_{{\bf k},n}^{GW}(\omega)$ and its derivatives would first lead to erroneous terms that are nonlinear in $t$ [\onlinecite{Mahan_book}] and, second, to miscounting of the correlated self-energy corrections. This introduces spurious features in the quasiparticle propagator (\ref{eq:Gcumul}) and its spectrum (\ref{eq:N}).

The only exception to the above counterexample is the contribution represented by the generic $G_{0}W_{0}$ diagram of Fig. \ref{Sigma_k} which is common to both the second order self-energy and cumulant series. Its two vertices are fully correlated through momentum conservation, and in the long time limit also through the energy conservation. For the present interaction $W$ there is no subtraction of uncorrelated processes from the second order cumulant and this leads to expressions (\ref{eq:Clarge}) and (\ref{eq:CsatGW}).  Therefore, if the second order cumulant gives the dominant contribution to the series (\ref{eq:Cl}), as is the case in weakly correlated multiexcitation processes, then the results of $G_{0}W_{0}$ calculations can be efficiently exploited in the calculations of threshold and satellite properties of the spectra (\ref{eq:N}). In contrast, the use of GW approximation with $G$ renormalized through some approximate or self-consistent form of self-energy $\Sigma_{{\bf k},n}^{GW}$ going beyond the $G_{0}W_{0}$-level would give rise to unreliable results for $G$ obtained from the cumulant representation (\ref{eq:Gcumul}).

Lastly, we note that in the calculations of optical response functions, i.e. the e-h pair propagators, the situation regarding the consistency of employing higher order GW corrections for the one particle propagators is different. Namely, here the inclusion of e-h vertex corrections calculated in the ladder approximation is consistent with the use of higher order GW self energies for the propagators [\onlinecite{MahanGW}], and this route has been followed in the calculations of optical properties of solids [\onlinecite{MariniDelSole,YAMBO,Bechstedt_vertex,Onida2,Attaccalite}].

\section{Protocol for implementation of the GW+C approach}
\label{sec:protocol}

In this section we summarize the results of this work in the form of a protocol for consistent implementation of {\it ab initio} GW and cumulant (GW+C) approach to the calculations of quasiparticle propagators and spectra. It encompasses several steps:

(i) The one-particle eigenstates $|{\bf k},n\rangle$ and eigenenergies $\epsilon_{{\bf k},n}^{0}$ that diagonalize the effective one particle Hamiltonian (\ref{eq:Hqp}) and define the band structure of the solid in the absence of screened interaction $W$ are obtained using some well established scheme. In practice these are most often the Kohn-Sham (KS) one particle states and energies calculated within the DFT employing a static (either local, semi-local or non-local) exchange and correlation potential $v_{xc}({\bf r})$. 

(ii) Using the one particle states and energies obtained in (i) the dynamic electronic response function $\chi_{\bf q}(\omega')$ of the system is calculated within the RPA or one of its improvements. From this one determines the excitation spectrum ${\cal D}_{\bf q}(\omega')$ needed in the calculation of the $G_{0}W_{0}$ quasiparticle self-energy defined in (\ref{eq:SigmaGW})-(\ref{eq:Sigma2k}). 

(iii) Following the arguments presented in Secs. \ref{sec:propagators_cumul} and \ref{sec:L}, and summarized in Sec. \ref{sec:GWlink}, the only justifiable consistent connection of the {\it ab initio} GW input and cumulant expansion can be established on the level of the $G_{0}W_{0}$ self-energy combined with the second order cumulant. Otherwise, uncontrolable overcounting effects may be encounterred. This GW+C procedure yields for quasiparticles in the Bloch states in the $n$-th band 
\barr
C_{{\bf k},n}^{\stackrel{>}{\scriptscriptstyle <}}(t)&=& -\int_{\stackrel{\mu}{\scriptscriptstyle -\infty}}^{\stackrel{\infty}{\scriptscriptstyle \mu}} d\omega' \frac{\mp \frac{1}{\pi}\mbox{Im}\Sigma_{{\bf k},n}^{\stackrel{>}{\scriptscriptstyle <}}(\omega')}{(\omega'-\epsilon_{{\bf k},n}^{0})^2}\left[1-e^{-i(\omega'-\epsilon_{{\bf k},n}^{0})t}\right]\nonumber\\
&-&
 i\left(\mbox{Re}\Sigma_{{\bf k},n}^{\stackrel{>}{\scriptscriptstyle <}}(\epsilon_{{\bf k},n}^{0})-i v_{{\bf k},n}^{xc}\right)t. 
\label{eq:CSigma}
\earr
Here the upper and lower symbols refer to electron and hole, respectively, and $\Sigma_{{\bf k},n}^{\stackrel{>}{\scriptscriptstyle <}}(\omega')$ is the full $G_{0}W_{0}$ quasiparticle self-energy satisfying $\mp\mbox{Im}\Sigma_{{\bf k},n}^{\stackrel{>}{\scriptscriptstyle <}}(\omega')\geq 0$. The second term on the RHS of (\ref{eq:CSigma}) follows from Kramers-Kronig relations applied to the linear $t$-term of $C_{{\bf k},n}^{\stackrel{>}{\scriptscriptstyle <}}(t)$ from which $v_{{\bf k},n}^{xc}$ of Eq. (\ref{eq:C1}) is subtracted if the unperturbed $|{\bf k},n\rangle$ and $\epsilon_{{\bf k},n}^{0}$ are the KS states and eigenenergies, respectively. Calculations of the ultrafast phenomena involving quasiparticle propagators should use this {\it nonasymptotic} form of the cumulants [\onlinecite{SGLD,Lazic_PRL,Lazic_PRB}]. If the first term on the RHS of (\ref{eq:CSigma}) is treated separately from the other terms the factor $\pm i\eta$ may be added {\it ad hoc} into the denominator in order to remove spurious divergences from the expressions for satellite spectra.

(iv) Calculations of the quasiparticle spectra for comparison with the results of steady-state measurements (e.g. cw-photoemission) may use the simpler long time limit of (\ref{eq:CSigma}) based on the ansatz used in (\ref{eq:limC}). This enables carrying out the analytical treatment much farther and leads to the spectrum (\ref{eq:Nsat1}) whose practical implementation is demonstrated in Sec. \ref{sec:Si}.

\section{Spectral function of silicon valence bands in the GW+C approach}
\label{sec:Si}
 

\begin{figure*}[t]
\includegraphics[width=0.95\textwidth]{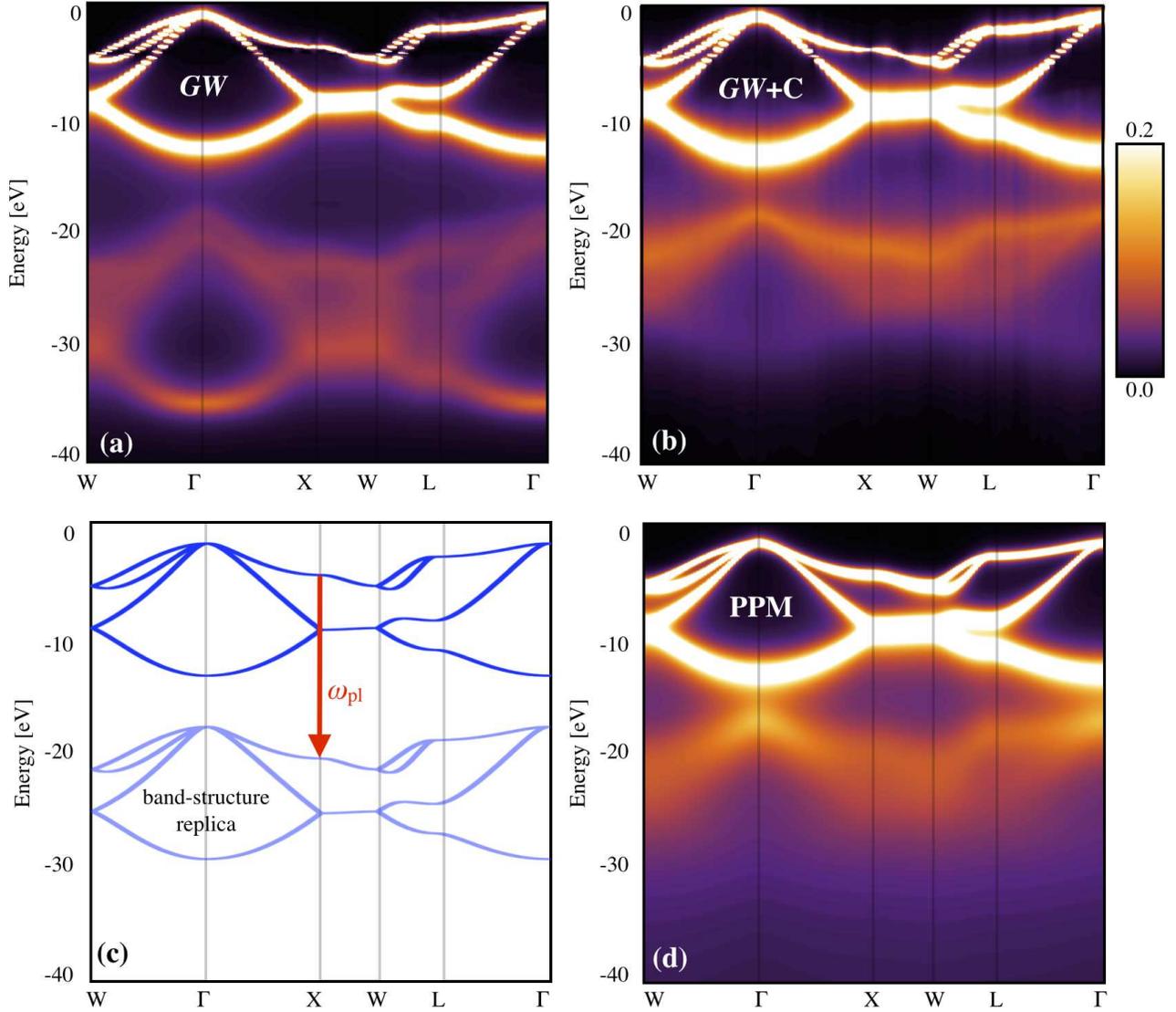}
\caption{(Color online)
Spectral function of silicon [in eV$^{-1}$ units] evaluated using (a) the $G_0W_0$ approximation, (b) the $G_0W_0$ + cumulant ($G_0W_0$+C) approach, 
(c) DFT-LDA, and (d) the plasmonic polaron model (PPM). For comparison, we report in panel (c) a replica of the DFT-LDA band structure
red-shifted by the plasmon energy $\omega_{\rm pl}$ ($\omega_{pl}\simeq16.6$~eV for silicon). All energies are relative to the Fermi level.}
\label{fig:Siall}
\end{figure*}

\begin{figure}[t]
\includegraphics[width=0.38\textwidth]{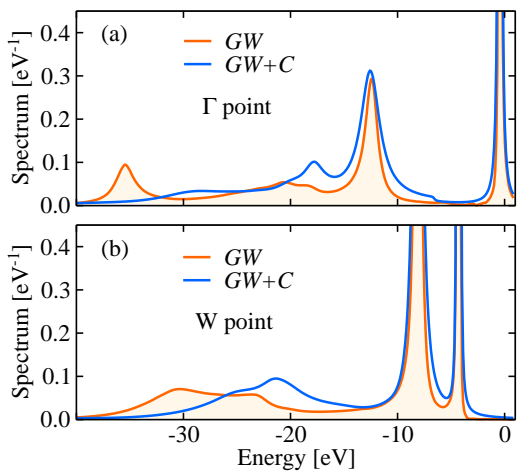}
\caption{(Color online)
Spectral function of silicon at the $\Gamma$ (a) and W (b) high symmetry points 
evaluated using the Sternheimer-$GW$ method ($G_0W_0$ approximation) and the $G_0W_0$ 
+ cumulant ($GW$+C) approach.}
\label{fig:spectrumG}
\end{figure}

We now proceed to discuss the application of cumulant expansion 
in the context of {\it ab initio} calculations of spectral functions of solids. 
In particular, we study the spectral properties of silicon and the 
signatures of electron-plasmon interaction based on three different approaches:  the $G_0W_0$ approximation, 
the $G_0W_0+C$ approach, and the plasmonic polaron model [\onlinecite{Caruso}].

In the $G_0W_0$ approximation the spectral function is evaluated as:
 \bq
\label{eq.specfun}
  {\cal N}_{\bf k}(\omega)  
  = \frac{1}{\pi}  \sum_n \frac{\Sigma''_{{\bf k},n} (\omega)}
  {[\omega - \epsilon_{{\bf k},n} -\Delta \Sigma'_{{\bf k},n}(\omega)]^2 
  + [\Sigma''_{{\bf k},n}(\omega)]^2},
  \eq
where $\Sigma'$ and $\Sigma''$ are the real and imaginary parts of the 
$G_0W_0$ self-energy, which we evaluate within the Sternheimer-$GW$ approach [\onlinecite{Giustino2,Lambert}].
$\epsilon_{{\bf k},n}^{0}$ denote Kohn-Sham density-functional theory (DFT) [\onlinecite{Hohenberg,Kohn}] eigenvalues in the $n$-th band and 
$\Delta \Sigma'_{{\bf k},n}(\omega) \equiv \Sigma'_{{\bf k},n}(\omega) - v^{\rm xc}_{n {\bf k}}$, 
where $v^{\rm xc}_{{\bf k},n}$ is the exchange-correlation potential. 
%
The $G_0W_0$ + C spectral function has been obtained following the procedure leading to expression (\ref{eq:Nsat1}).
%

Finally, in the plasmonic polaron model the following approximations are introduced in order to circumvent the numerical cost of $G_0W_0$ calculations:
(i) the quasiparticle linewidth is assumed to increase quadratically with the energy difference from the Fermi energy; 
(ii) the quasiparticle correction to the DFT eigenvalues are ignored; 
(iii) the imaginary part of the self-energy is approximated through a simple Lorentzian model.
A detailed description of this procedure may be found elsewhere [\onlinecite{Caruso}].

Density-functional theory calculations [\onlinecite{Hohenberg,Kohn}]
in the Perdew-Zunger local density approximation (LDA) [\onlinecite{PerdewZunger}]
are performed using the Quantum-ESPRESSO code [\onlinecite{Giannozzi}].
We used a norm-conserving Troullier-Martins pseudopotential [\onlinecite{Troullier}]
and a 18~Ry kinetic energy cutoff
to describe the wave functions. 
The Brillouin zone is discretized on a 6$\times$6$\times$6 Monkhorst-Pack grid.
The dielectric matrix of silicon has been evaluated using a 10~Ry kinetic energy cutoff,
whereas we used a 18~Ry cutoff for the exchange part of the self-energy.
The frequency dependence of the screened Coulomb interaction 
was determined using a Pad\'e approximant fit to 100 points 
along the imaginary frequency axis on a uniform grid from 0 to 25 eV.
The Sternheimer-$GW$ self-energy has been evaluated
on a discrete frequency grid with a 25~meV spacing.
The high-symmetry lines are computed with a momentum
spacing of 0.025$\times 2\pi/a$, where $a=10.26$~bohr
is the lattice constant of silicon.

In Fig.~\ref{fig:Siall}~(a), we show the $G_0W_0$ spectral function of silicon.
At binding energies between 0 and 12~eV, the spectral function exhibits 
intense features which correspond to quasiparticle excitations in the absence of plasmons. 
These spectral features define the ordinary valence bands of silicon. 
At binding energies larger than 18~eV, the spectral function reveals 
additional spectral features with a well defined dispersion
that resembles that of the quasiparticle bands. 
These spectral features are the band structures of plasmonic polarons [\onlinecite{Feliciano,Caruso}]
and stem from the simultaneous excitation of a hole and a plasmon.
In the $G_0W_0$ approximation, plasmonic polarons bands are found at a binding energy of approximately  
$1.2-1.5\omega_{\rm pl}$ below the quasiparticle bands (in silicon $\omega_{\rm pl}\simeq16.6$~eV). 
Overall, the $G_0W_0$ approximation overestimates 
the energy of these spectral features, which are typically found 
at $\sim\omega_{\rm pl}$ below the quasiparticle bands in photoemission 
experiment [\onlinecite{Guzzo,Fadley}].

As compared to $G_0W_0$ calculations, the quasiparticle bands are left essentially 
unaffected by the $G_0W_0$ + C approach, as illustrated in 
Fig.~\ref{fig:Siall}~(b) and Fig.~\ref{fig:spectrumG}.
On the other hand, the plasmonic polarons bands are red-shifted by $\sim\omega_{\rm pl}$
with respect to the quasiparticle bands, and are thus compatible 
with the energy range of plasmon satellite observed in XPS measurements [\onlinecite{Guzzo}].
To further emphasize the differences between the $G_0W_0$ and $G_0W_0$ + C spectral 
function we compare in Fig.~\ref{fig:spectrumG} the spectral function of silicon at 
the $\Gamma$ and W high symmetry points. 
Overall, the plasmonic polaron bands appear as a broadened low-intensity 
replica of the quasiparticle band structure, red-shifted by the plasmon energy $\omega_{\rm pl}$. 
The prediction of plasmonic polaron band structures [\onlinecite{Feliciano}] was subsequently
confirmed by angle-resolved photoemission spectroscopy experiments [\onlinecite{Fadley}].

The spectral function obtained from the plasmonic polaron model [Fig.~\ref{fig:Siall}~(d)] 
reproduces the main qualitative features of the $G_0W_0$ + C approach at a very small
computational cost. As compared to $G_0W_0$ + C calculations, the plasmonic polaron 
bands appear slightly more smeared out as a consequence of quadratic dependence of the 
linewidth on the binding energy. 
For comparison, Fig.~\ref{fig:Siall}~(c) reports the DFT-LDA band structure and a replica 
of the full valence band structure red-shifted by the plasmon energy $\omega_{\rm pl}$.

\section{Summary}
\label{sec:summary}

It is argued that under the conditions of weakly correlated successive scattering events the propagator of a quantum particle (single electron or hole) in interaction with bosonized excitations of the heatbath of a solid (single and collective electronic excitations, phonons, etc.) is obtained to a high level of accuracy from the second order cumulant expansion. The second order cumulant, which is proven to encompass the relevant dynamic features of the complete cumulant series (\ref{eq:Cl}), can be conveniently calculated from the corresponding $G_{0}W_{0}$-approximation expression for the quasiparticle self-energy [\onlinecite{MahanGW,QF1958,AHK}]. The direct connection is established via Eqs. (\ref{eq:rho2>}) and (\ref{eq:rho2<}) which yield the $G_{0}W_{0}$-derived joint densities of excitations required to calculate the second order cumulant (\ref{eq:C_rho2}).  It is also emphasized that the use of higher order forms of the GW approximation is not justified for the second and higher order cumulants. Thus, it is recommended that cmulant expansion be used in combination with the $G_0W_0$ approximation. Using silicon as a test case we have illustrated the modifications of the quasiparticle spectral function which one obtains when moving from the standard $G_0W_0$ approximation to the more sophisticated cumulant expansion. We point out that the calculations are in good agreement with recent photoemission experiments [\onlinecite{Fadley}]. 

The approach developed in the present work is rigorous in the limit of a quasiparticle propagating far above or below the Fermi level. However, the point at which it breaks down  is system specific and could be estimated only by carrying out the standard Feynman-Dyson expansion of the same Green's function which enables the treatment of quasiparticle propagation in the vicinity of the Fermi level in both time directions (cf. discussion in Sec. III of Ref. [\onlinecite{Doniach}] and in the last paragraph of Sec. \ref{sec:validityC2} above). 
In this context, it appears that the formulation of Ref. [\onlinecite{Kas}] would interpolate smoothly between the $G_{0}W_{0}$ limits of our expressions for single electron and hole propagators
(see also Sec. B of the Supplemental material [\onlinecite{Suppl}]); therefore, the present formulation and that of Ref. [\onlinecite{Kas}] should provide equivalent descriptions for quasiparticles far
from the chemical potential. The equivalence of the two approaches close to the chemical potential is a more complex question because, as pointed out in Sec. \ref{sec:validityC2}, in this case the second order cumulant
expansion breaks down, and the standard Feynman-Dyson perturbation expansion of the self-energy becomes more accurate [\onlinecite{Doniach}].
 We hope that this manuscript will stimulate further work on cumulant expansion and its applications to the studies of dynamical properties of quasiparticles in real materials and to the calculations of photoemission spectra.

\acknowledgments

V.K. acknowledges partial support by the Croatian 
Science Foundation under the project 3526.
F.C., H.L. and F.G. acknowledge support from the 
Leverhulme Trust (Grant RL-2012-001), the UK Engineering and Physical
Sciences Research Council (Grant No. EP/J009857/1), the University of Oxford Advanced Research 
Computing (ARC) facility, and the ARCHER UK National Supercomputing Service.




\begin{thebibliography}{11}
\bibitem{correlations} That such additional small expansion parameter may exist in the polaron problem was anticipated in Ref. [\onlinecite{vanHaeringen}].
\end{thebibliography}
\end{document}


\title{SUPPLEMENTAL MATERIAL \\ On the combined use of GW approximation and cumulant expansion in the calculations of quasiparticle spectra: The paradigm of Si valence bands}
\author{Branko Gumhalter\footnote{Corresponding author. Email: branko@ifs.hr}}
\affiliation{Institute of Physics, HR-10000 Zagreb, Croatia}
\author{Vjekoslav Kova\v{c}}
\affiliation{Department of Mathematics, University of Zagreb, HR-10000 Zagreb, Croatia}
\author{Fabio Caruso}
\affiliation{Department of Materials, University of Oxford, Parks Road, Oxford OX1 3PH, United Kingdom}
\author{Henry Lambert}
\affiliation{Department of Materials, University of Oxford, Parks Road, Oxford OX1 3PH, United Kingdom}
\author{Feliciano Giustino}
\affiliation{Department of Materials, University of Oxford, Parks Road, Oxford OX1 3PH, United Kingdom}


\pacs{
71.10.-w,    
71.15.-m,     
71.15.Qe,    
71.45.Gm    
}

\maketitle

\newcommand{\bq}{\begin{equation}}
\newcommand{\eq}{\end{equation}}

\newcommand{\barr}{\begin{eqnarray}}
\newcommand{\earr}{\end{eqnarray}}

\appendix
\section{Integral representation of cumulant series}
\label{sec:A1}

The purpose of this appendix is to provide a rigorous mathematical foundation for the transformation (37) and its inverse (39) of the main text. We omit subscripts and superscripts and write simply $\rho$ and $C$. Our goal is to define the mapping $\rho(\nu)\mapsto C(t)$, specify its domain and range, prove that it is one-to-one and onto, and describe its inverse map $C(t)\mapsto \rho(\nu)$. Moreover, at the end of the section we restrict our attention to a rather generic particular case that is sufficient for the application explained in the main part of this paper.

Let $\mathcal{M}^+\!(\mathbb{R})$ and $\mathcal{M}(\mathbb{R})$ respectively denote the space of \emph{finite positive measures} and the space of \emph{finite signed measures} on the real line (cf. Sections 1.2 and 4.1 in [\onlinecite{Cohn_book}]). Indeed, an element of $\mathcal{M}(\mathbb{R})$ is simply a difference of two elements from $\mathcal{M}^+\!(\mathbb{R})$. This space contains all absolutely integrable real-valued functions, but it also contains the ``true'' measures, such as absolutely summable linear combinations of Dirac delta-functions at certain points of $\mathbb{R}$. For any finite signed measure $\rho\in \mathcal{M}(\mathbb{R})$ we interpret (37)of the main text as
\bq
C(t) = \int_{-\infty}^{\infty} \frac{e^{-i\nu t}-1+i\nu t}{\nu^2} \rho(\nu) d\nu.
\label{eq:C_drho2}
\eq
This transformation is well-defined since the function $(e^{-i\nu t}-1+i\nu t)/\nu^2$ can be extended at $\nu=0$ in a way that it becomes a bounded continuous function on the whole real line and thus it can be integrated against the measure $\rho$. The basics of integration theory with respect to abstract measures can be found in the classical textbooks on measure theory, such as [\onlinecite{Cohn_book}] and [\onlinecite{Folland_book}]. Let us remark that integration with respect to a general measure $\rho$ is usually written with $\rho(\nu) d\nu$ replaced by $d\rho(\nu)$ or $\rho(d\nu)$ in mathematical literature in order to emphasize that $\rho$ need not have a density (i.e.\@ it could be a true measure). We will stick to the former notation, since it is also common to understand signed (or even complex) measures as particular cases of distributions (cf. Section 6.11 in [\onlinecite{Rudin_book}]).

Differentiation of (\ref{eq:C_drho2}) is justified using the dominated convergence theorem (cf. Theorem~2.27(b) from [\onlinecite{Folland_book}]) and it yields
\barr
\dot{C}(t) = -i \int_{-\infty}^{\infty} \frac{e^{-i\nu t}-1}{\nu} \rho(\nu) d\nu,
\label{eq:C_rhoder}
\\
\ddot{C}(t) = -\int_{-\infty}^{\infty} e^{-i\nu t} \rho(\nu) d\nu.
\label{eq:C_rho2der}
\earr
Therefore $\ddot{C}=-\widehat{\rho}$, where $\widehat{\rho}$ denotes the so-called \emph{Fourier-Stieltjes transform} of $\rho$ (see Chapter VI of [\onlinecite{Katznelson_book}]), also known simply as the Fourier transform of a measure $\rho$ (see Chapter IX of [\onlinecite{ReedSimon2_book}] or Chapter 4 in [\onlinecite{Folland2_book}]), which is a function $\widehat{\rho}\colon\mathbb{R}\to\mathbb{C}$ defined by
\bq
\widehat{\rho}(t) = \int_{-\infty}^{\infty} e^{-i\nu t} \rho(\nu) d\nu.
\label{eq:fstransform}
\eq
In the literature on probability theory (such as Chapter 3 of [\onlinecite{Durrett_book}]), $\widehat{\rho}$ is usually called the characteristic function of $\rho$, even though the normalization is slightly different there. Substituting $t=0$ into (\ref{eq:C_drho2}) and (\ref{eq:C_rhoder}) gives
\barr
C(0)=0,
\label{eq:unitarity2}
\\
\dot{C}(0)=0.
\label{eq:zerowork2}
\earr
From (\ref{eq:C_rho2der})--(\ref{eq:zerowork2}) we see that an equivalent way of expressing $C$ in terms of $\rho$ is
\bq
C(t) = \int_{0}^{t} (s-t) \widehat{\rho}(s) ds,
\eq
with the usual convention that $\int_{0}^{t}=-\int_{t}^{0}$ when $t<0$. It is easy to see that $C(t)$ is always two times continuously differentiable, but it is not true that every such function can be obtained from some $\rho\in \mathcal{M}(\mathbb{R})$, as we shall soon see.

Now we claim that the transform defined by (\ref{eq:C_drho2}) is a one-to-one map from $\mathcal{M}(\mathbb{R})$ to some collection of twice continuously differentiable functions, i.e.\@ that different measures $\rho$ map to different functions $C$. In order to verify that we assume $\rho_1\mapsto C$ and $\rho_2\mapsto C$. From (\ref{eq:C_rho2der}) we get
\bq
\widehat{\rho}_1 = -\ddot{C} = \widehat{\rho}_2,
\eq
so the well-known fact that the Fourier-Stieltjes transform $\rho\mapsto\widehat{\rho}$ is one-to-one (see Subsection VI.2.2 in [\onlinecite{Katznelson_book}] or Theorem 4.33 in [\onlinecite{Folland2_book}]) gives $\rho_1=\rho_2$, as desired.

Next, let us clarify the image of the transform $\rho\mapsto C$ defined by (\ref{eq:C_drho2}). A highly nontrivial result we need is the characterization of the images of $\mathcal{M}^+\!(\mathbb{R})$ and $\mathcal{M}(\mathbb{R})$ with respect to the assignment $\rho\mapsto\widehat{\rho}$. \emph{Bochner's theorem} (see Subsection VI.2.8 in [\onlinecite{Katznelson_book}], Theorem IX.9 in [\onlinecite{ReedSimon2_book}], or Theorem 4.18 in [\onlinecite{Folland2_book}]) states that the Fourier-Stieltjes transforms $\widehat{\rho}$ of finite positive measures $\rho$ are precisely the continuous functions $B\colon\mathbb{R}\to\mathbb{C}$ satisfying
\bq
\sum_{j,k=1}^{n} B(t_j-t_k) \xi_j \overline{\xi_k} \geq 0
\label{eq:positivetype}
\eq
for any choice of points $t_1,\ldots,t_n\in\mathbb{R}$ and any choice of numbers $\xi_1,\ldots,\xi_n\in\mathbb{C}$. Functions $B$ with property (\ref{eq:positivetype}) are called \emph{functions of positive type} or \emph{positive definite functions}. If $\mathcal{P}^+\!(\mathbb{R})$ denotes the collection of all continuous functions of positive type, then by Bochner's theorem we also know that the image of $\mathcal{M}(\mathbb{R})$ over $\rho\mapsto\widehat{\rho}$ is the collection $\mathcal{P}(\mathbb{R})$ of all functions that can be written as a difference of two functions from $\mathcal{P}^+\!(\mathbb{R})$. Therefore, our transformation $\rho(\nu)\mapsto C(t)$ is a one-to-one and onto map between the space of finite signed measures $\mathcal{M}(\mathbb{R})$ and the collection of all twice differentiable functions $C(t)$ such that $\ddot{C}(t)$ belongs to the space $\mathcal{P}(\mathbb{R})$. We used the word ``space'' in connection with $\mathcal{M}(\mathbb{R})$ and $\mathcal{P}(\mathbb{R})$ because these are indeed vector spaces on which appropriate norms can be defined; cf. Refs. [\onlinecite{Cohn_book}] or [\onlinecite{Folland_book}].

What remains is the question of retrieving $\rho$ back from $C$, which in turn reduces to the problem of recovering a measure $\rho$ from its Fourier-Stieltjes transform $\widehat{\rho}$. Technical complications come from the fact that Formula (39) of the main text can only be applied when $\rho$ is a proper function (i.e.\@ not a true measure) and $\ddot{C}(t)$ is absolutely integrable. The techniques for inverting the assignment $\rho\mapsto\widehat{\rho}$ are known as the \emph{summation methods for Fourier integrals} in the literature on classical harmonic analysis and are equivalent to retrieving a distribution from its characteristic function in probability theory. Completely generally, for any two real numbers $a<b$ we have
\bq
\rho(a,b) + \frac{1}{2}\rho(\{a,b\}) = \lim_{T\to\infty} \frac{1}{2\pi}\int_{-T}^{T} \frac{e^{ibt}-e^{iat}}{it} \widehat{\rho}(t) dt
\label{eq:inversefs}
\eq
(see Theorem~3.3.4 in [\onlinecite{Durrett_book}]) and
\bq
\rho(\{a\}) = \lim_{T\to\infty} \frac{1}{2T} \int_{-T}^{T} e^{iat} \widehat{\rho}(t) dt.
\label{eq:inversefs2}
\eq
Here $\rho(a,b)$ denotes the number that the measure $\rho$ assigns to an interval $(a,b)$, which can be alternatively written as $\int_{(a,b)}\rho(\nu)d\nu$, while $\rho(\{a\})$ is the ``mass'' of the measure $\rho$ at a point $\nu=a$. In (\ref{eq:inversefs}) we write $\lim_{T\to\infty}\int_{-T}^{T}$ instead of simply $\int_{-\infty}^{\infty}$ in order to emphasize that the above integral is not necessarily absolutely convergent and that it is taken in the ``principal value'' sense. Equations (\ref{eq:inversefs}) and (\ref{eq:inversefs2}) finally allow us to write
\begin{widetext}
\bq
\rho(a,b) = -\lim_{T\to\infty} \frac{1}{2\pi}\int_{-T}^{T} \frac{e^{ibt}-e^{iat}}{it} \ddot{C}(t) dt
+ \lim_{T\to\infty} \frac{1}{2T} \int_{-T}^{T} \frac{e^{iat}+e^{ibt}}{2} \ddot{C}(t) dt.
\eq
\end{widetext}
That way the measure $\rho$ is determined on all bounded open intervals $(a,b)$, which specifies it completely. If $\rho$ is simply an integrable function, then we have more straightforward ways of recovering $\rho(\nu)$ pointwise, such as
\bq
\rho(\nu) = -\lim_{T\to\infty} \frac{1}{2\pi}\int_{-T}^{T} \Big(1-\frac{|t|}{T}\Big) e^{i\nu t} \ddot{C}(t) dt
\label{eq:Fourinver1}
\eq
(cf. Subsection VI.1.11 in [\onlinecite{Katznelson_book}] or Theorem 8.35 in [\onlinecite{Folland_book}]). Additionally, if $\widehat{\rho}$ is also integrable, then we have the simplest inversion formula,
\bq
\rho(\nu) = -\frac{1}{2\pi}\int_{-\infty}^{\infty} e^{i\nu t} \ddot{C}(t) dt
\label{eq:Fourinver2}
\eq
(cf. Subsection VI.1.12 in [\onlinecite{Katznelson_book}] or Theorem 8.26 in [\onlinecite{Folland_book}]). If $\rho$ is not assumed to be continuous, then formulae (\ref{eq:Fourinver1}) and (\ref{eq:Fourinver2}) do not necessarily hold for each $\nu\in\mathbb{R}$; they can fail on a set of measure $0$, which is considered negligible. Let us also comment that the assignment $C\mapsto\rho$ (as opposed to $\rho\mapsto C$) is natural even at the level of the so-called \emph{tempered distributions}, because both differentiation and the inverse Fourier transform make sense there. That viewpoint is too abstract to be useful for us here, so we only refer to Section V.3 of [\onlinecite{ReedSimon1_book}] or Chapter 6 of [\onlinecite{Rudin_book}].

For a reader who is primarily interested in an operatively computational side of the correspondence $\rho(\nu)\leftrightarrow C(t)$, let us discuss a particular case when a finite signed measure $\rho\in \mathcal{M}(\mathbb{R})$ consists of an absolutely continuous part with density $f$ and an absolutely convergent series of delta-functions (cf. Eq. (58) of the main text). More precisely, we assume that
\bq
\rho(\nu) = f(\nu) + \sum_{j} \alpha_j \delta(\nu-\nu_j),
\eq
where $f$ is a continuous function satisfying $\int_{-\infty}^{\infty}|f(\nu)|d\nu<\infty$, we take a sequence of positive numbers $\nu_1<\nu_2<\nu_3<\cdots$ without a finite limit, and we choose a sequence of real scalars $\alpha_j$ such that $\sum_{j}|\alpha_j|<\infty$. Let us emphasize that we impose no conditions on the finiteness of any moments of $\rho$. Definition (\ref{eq:C_drho2}) now reads
%
\barr
C(t)&=& \int_{-\infty}^{\infty} \frac{e^{-i\nu t}-1+i\nu t}{\nu^2} f(\nu) d\nu\nonumber\\
&+&
\sum_{j} \alpha_j \frac{e^{-i\nu_j t}-1+i\nu_j t}{\nu_j^2}.
\earr
%
Conversely, if we want to recover $\rho$ from $C$, we first apply (\ref{eq:inversefs2}) to get
\bq
\alpha_j = -\lim_{T\to\infty} \frac{1}{2T} \int_{-T}^{T} e^{i\nu_j t} \ddot{C}(t) dt.
\eq
The points $\nu_j$ with $\alpha_j\neq 0$ are identified at the same time as the only points $\nu\in\mathbb{R}$ for which we have
\bq
\lim_{T\to\infty} \frac{1}{2T} \int_{-T}^{T} e^{i\nu t} \ddot{C}(t) dt \neq 0.
\eq
Now we turn our attention to the absolutely continuous part (i.e.\@ the part corresponding to the density function $f$),
$$ C_{\mathrm{ac}}(t) = C(t) - \sum_{j} \alpha_j \frac{e^{-i\nu_j t}-1+i\nu_j t}{\nu_j^2}. $$
Using (\ref{eq:Fourinver1}) we finally obtain
\bq
f(\nu) = -\lim_{T\to\infty} \frac{1}{2\pi}\int_{-T}^{T} \Big(1-\frac{|t|}{T}\Big) e^{i\nu t} \ddot{C}_{\mathrm{ac}}(t) dt.
\eq
Once again, if $|\ddot{C}_{\mathrm{ac}}(t)|$ has a finite integral over the whole real line, then we can apply the simpler formula (\ref{eq:Fourinver2}), which gives
\bq
f(\nu) = -\frac{1}{2\pi}\int_{-\infty}^{\infty} e^{i\nu t} \ddot{C}_{\mathrm{ac}}(t) dt.
\eq

\section{Drag boson representation of the cumulant series}
\label{sec:Drag_Boson}

Bijective integral representation (37) of the main text holds for the case of cumulant representation of both the single electron and single hole propagators in empty and occupied bands that constitute expression (5) of the main text, respectively. Thus, the cumulant encompassing the electron (e) and hole (h) cases can be formally represented as 
\barr
C_{\bf k}(t)&=& C_{\bf k}^{\rm e}(t)+C_{\bf k}^{\rm h}(t)\nonumber\\
 &=&
 -\theta(t)(1-n_{\bf k})\int d\nu \rho_{\bf k}^{>}(\nu)\frac{1-i\nu t-e^{-i\nu t}}{\nu^2} \nonumber\\
& &
-\theta(-t)n_{\bf k}\int d\nu \rho_{\bf k}^{<}(\nu)\frac{1+i\nu t-e^{i\nu t}}{\nu^2}.
\label{eq:C><}
\earr
Here $n_{\bf k}$ has the same meaning as in (7) and (8) of the main text, viz. it is the electron occupation number of the initial state ${\bf k}$. The excitation densities  $\rho_{\bf k}^{>}(\nu)$ and $\rho_{\bf k}^{<}(\nu)$ involve the one-particle states only above and below the Fermi level, respectively, and hence are inequivalent (cf. the lowest order terms (34) and (35) of the main text). $C_{\bf k}^{\rm e}(t)$ and $C_{\bf k}^{\rm h}(t)$ are obtained from each other by the replacements $\rho_{\bf k}^{>}(\nu)\leftrightarrow\rho_{\bf k}^{<}(\nu)$, $(1-n_{\bf k})\leftrightarrow n_{\bf k}$, and time reversal. 
Following this a close connection can be made between the formally time ordered cumulant (\ref{eq:C><}) taken in the $G_{0}W_{0}$ limit and the $G_{0}W_{0}$ representation of the retarded cumulant given by expression (7) of Ref. [\onlinecite{KasSuppl}]. Assuming a single particle propagation, and in both cases the averaging over the $W$-noninteracting ground state in the definition of Green's function (5) of the main text,  the two cumulants become equivalent through time reversal of the hole component in (\ref{eq:C><}).

Quite formally, the unified representation of cumulant series (\ref{eq:C><}) appropriate to {\it diagonal} single particle propagators deriving from Hamiltonian (1) of the main text can be obtained from the expression for the second order cumulant involving the unperturbed single particle propagator $G_{\bf k}^{0}$ given by (8) of the main text in {\it all} intermediate states (likewise the core hole case) and the associated {\it "drag boson"} (DB) propagator 
\barr
D_{\bf k}^{DB}(t)= &-& i\theta(t)(1-n_{\bf k})\int d\nu \rho_{\bf k}^{>}(\nu)e^{-i\nu t} \nonumber\\
&-&
i\theta(-t)n_{\bf k}\int d\nu \rho_{\bf k}^{<}(\nu)e^{i\nu t}.
\label{eq:DB}
\earr
Here $n_{\bf k}$ is a pure number (0 or 1) because in the present formulation it is conserved during the entire interval of propagation of a quasiparticle once the quantum number ${\bf k}$ in (\ref{eq:C><}) has been fixed. However, due to the absence of invariance $|\rho_{\bf k}(\nu)|\neq|\rho_{\bf k}(-\nu)|$ there is no full analogy between the dispersions of expression (\ref{eq:DB}) and the standard boson propagator (10) of the main text. The thus formulated drag boson is associated with the parent unperturbed ${\bf k}$-state of the electron ($>$) or the hole ($<$) with which it can exchange excitation energy $\nu$ but not the momentum since by construction the momenta ${\bf q}$ exchanged with the original bosons have been integrated over in $\rho_{\bf k}^{>}(\nu)$ and $\rho_{\bf k}^{<}(\nu)$. 
Hence, the manifestation of this interaction in (\ref{eq:C><}) is equivalent to that between  the recoilless parent quasiparticle and bosons that are dragged along during the entire intermediate propagation interval, likewise in the forced oscillator model [\onlinecite{SunkoAJP}].  A diagrammatic representation of this forward scattering process throughout the propagation interval is shown in supplementary Fig. \ref{DragBoson}. The representation of (\ref{eq:DB}) in terms of the $G_{0}W_{0}$ self-energy follows directly by making use of (34) and (35) of the main text.   

\begin{figure}[tb]
\rotatebox{0}{ \epsfxsize=9cm \epsffile{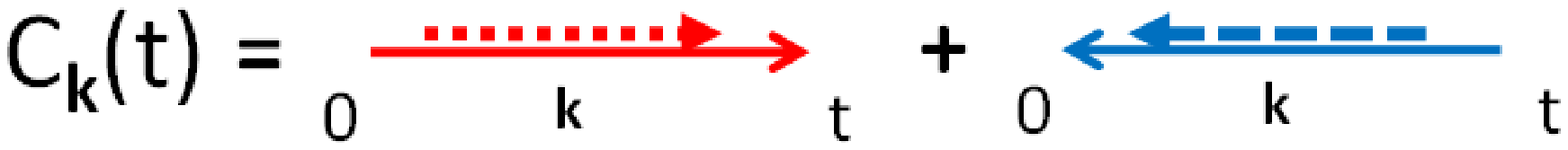} } 
\caption{ (Color online) Drag boson representation of the cumulant $C_{\bf k}(t)=C_{\bf k}^{\rm e}(t)+ C_{\bf k}^{\rm h}(t)$, Eq. (\ref{eq:C><}), for both directions of time propagation: $t>0$ for electron (e), and $t<0$ for hole (h). Full line denotes the propagator of unperturbed single electron (single hole) and the dashed line of the drag boson. Different dashings indicate the inequivalence of $\rho_{\bf k}^{>}(\nu)$ and $\rho_{\bf k}^{<}(\nu)$ in the drag boson propagators associated with electrons and holes, respectively.}
\label{DragBoson}
\end{figure}

To demostrate the construction of the DB propagator (\ref{eq:DB}) by field theoretical methods we introduce in the first step the drag boson creation and annihilation operators in the time domain for the parent electron ${\bf k}$-states above the Fermi energy (symbol $>$). We define the DB creation operator by
\bq
{\cal B}_{\bf k>}^{\dag}(t)=(1-n_{\bf k})\int d\nu e^{i\nu t}\rho_{\bf k}^{>}(\nu){\cal B}_{\bf k>}^{\dag}(\nu), 
\label{eq:B>+}
\eq
and the conjugated annihilation operator ${\cal B}_{\bf k>}(t)$ is obtained from (\ref{eq:B>+}) by hermitian conjugation. Likewise, for the parent hole states ${\bf k}$ below the Fermi energy (symbol $<$) we define 
\bq
{\cal B}_{\bf k<}^{\dag}(t)=n_{\bf k}\int d\nu e^{i\nu t}\rho_{\bf k}^{<}(\nu){\cal B}_{\bf k<}^{\dag}(\nu), 
\label{eq:B<+}
\eq
and ${\cal B}_{\bf k<}(t)$ is obtained from (\ref{eq:B<+}) by hermitian conjugation. In the energy domain the operators ${\cal B}_{\bf k>}(\nu)$ and ${\cal B}_{\bf k>}^{\dag}(\nu)$ satisfy the commutation relations 
\bq
\left[{\cal B}_{{\bf k}>}(\nu),{\cal B}_{{\bf k'}>}^{\dag}(\nu')\right]=\frac{\delta_{\bf k,k'}\delta(\nu-\nu')}{\rho_{\bf k}^{>}(\nu)},
\label{eq:DBcommutator}
\eq
and analogously so for the states ${\bf k}$ below the Fermi energy (symbol $<$). 

In the second step we exploit the assumption of the single particle propagation in the system and introduce the ${\bf k}$-restricted time ordering operator $T_{\bf k}$ which associates the electron ${\bf k}$-states above the Fermi energy only with the positive time propagation of drag bosons, and the hole ${\bf k}$-states below the Fermi energy only with negative time propagation of drag bosons. Hence, its action on a product of two drag boson operators ${\cal X}_{\bf k}(t)$ and ${\cal Y}_{\bf k}(t')$ is defined by 
\barr
T_{\bf k}\left[{\cal X}_{\bf k}(t){\cal Y}_{\bf k}(t')\right]&=& {\cal X}_{\bf k}(t){\cal Y}_{\bf k}(t')\theta(t-t')(1-n_{\bf k})\nonumber\\
&+&
{\cal Y}_{\bf k}(t'){\cal X}_{\bf k}(t)\theta(t'-t)n_{\bf k}.
\label{eq:Tk}
\earr
Hence, defining the drag boson propagator associated with the parent quasiparticle state ${\bf k}$ by
\bq
-i\langle 0|T_{\bf k}\left[\left({\cal B}_{\bf k}(t)+{\cal B}_{\bf k}^{\dag}(t)\right)\left({\cal B}_{\bf k}(t')+{\cal B}_{\bf k}^{\dag}(t')\right)\right]|0\rangle, 
\label{eq:D_DB}
\eq
we find that it is equal to expression (\ref{eq:DB}) which in combination with $G_{\bf k}^{0}$ produces the second order cumulant of the form (\ref{eq:C><}). 

Using the above definitions we can construct the effective Hamiltonian describing quasiparticle forward scattering by drag bosons 
\barr
H^{DB}&=& \sum_{\bf k}\epsilon_{\bf k}^{0}c_{\bf k}^{\dag}c_{\bf k} +\sum_{\bf k>} \int d\nu\rho_{\bf k}^{>}(\nu)\nu{\cal B}_{{\bf k}>}^{\dag}(\nu){\cal B}_{{\bf k}>}(\nu) \nonumber\\
&+&
\sum_{\bf k<} \int d\nu\rho_{\bf k}^{<}(\nu)\nu{\cal B}_{{\bf k}<}^{\dag}(\nu){\cal B}_{{\bf k}<}(\nu) \nonumber\\
&+&
\sum_{{\bf k}>}c_{\bf k}^{\dag}c_{\bf k}\int d\nu\rho_{\bf k}^{>}(\nu)\left({\cal B}_{{\bf k}>}(\nu)+{\cal B}_{{\bf k}>}^{\dag}(\nu)\right)\nonumber\\
&-& 
\sum_{{\bf k}<}c_{\bf k}c_{\bf k}^{\dag}\int d\nu \rho_{\bf k}^{<}(\nu)\left({\cal B}_{{\bf k}<}(\nu)+{\cal B}_{{\bf k}<}^{\dag}(\nu)\right). 
\label{eq:VDB}
\earr
The forward scattering form of the last two terms on the RHS of (\ref{eq:VDB}) inherently implies the second order cumulant solution (\ref{eq:C><}) for the diagonal quasiparticle propagators defined by the the original model Hamiltonian (1) of the main text. However, let us reiterate that it is not applicable to the studies of processes described by the offdiagonal elements of the Green's functions, scattering matrices etc, defined within the original model.

